\documentclass[reprint,aps,superscriptaddress]{revtex4-2}
\usepackage{bm}
\usepackage{amsmath}
\usepackage{amssymb}
\usepackage{graphicx}
\usepackage[export]{adjustbox}
\usepackage[colorlinks=true,linkcolor=blue,citecolor=blue,urlcolor=blue]{hyperref}
\usepackage{siunitx}
\usepackage{color}
\usepackage{braket}
\usepackage{soul}

\begin{document}

\title{Gate-tunable topological phases in superlattice modulated bilayer graphene}
\author{Yongxin Zeng}
\affiliation{Department of Physics, Columbia University, New York, NY 10027}
\affiliation{Center for Computational Quantum Physics, Flatiron Institute, New York, NY 10010}
\affiliation{Department of Physics, University of Texas at Austin, Austin, TX 78712}
\author{Tobias M. R. Wolf}
\affiliation{Department of Physics, University of Texas at Austin, Austin, TX 78712}
\author{Chunli Huang}
\affiliation{Department of Physics and Astronomy, University of Kentucky, Lexington, Kentucky 40506-0055}
\author{Nemin Wei}
\affiliation{Department of Physics, Yale University, New Haven, CT 06520}
\affiliation{Department of Physics, University of Texas at Austin, Austin, TX 78712}
\author{Sayed Ali Akbar Ghorashi}
\affiliation{Department of Physics and Astronomy, Stony Brook University, Stony Brook, NY 11794}
\author{Allan H. MacDonald}
\affiliation{Department of Physics, University of Texas at Austin, Austin, TX 78712}
\author{Jennifer Cano}
\affiliation{Department of Physics and Astronomy, Stony Brook University, Stony Brook, NY 11794}
\affiliation{Center for Computational Quantum Physics, Flatiron Institute, New York, NY 10010}

\date{\today}

\begin{abstract}
Superlattice potential modulation can produce flat minibands in Bernal-stacked bilayer graphene. In this work we study how band topology
and interaction-induced symmetry-broken phases in this system are controlled by tuning the displacement 
field and the shape and strength of the superlattice potential.
We use an analytic perturbative analysis to demonstrate that 
topological flat bands are favored by a honeycomb-lattice-shaped potential, and numerics to 
show that the robustness of topological bands depends on both the displacement field strength and the periodicity of the superlattice potential.  At integer fillings of the topological flat bands, 
the strength of the displacement field and the superlattice potential 
tune phase transitions between quantum anomalous Hall insulator, trivial insulator, and metallic states.
We present mean-field phase diagrams in a gate voltage parameter space at filling factor $\nu=1$, and discuss the prospects of realizing quantum anomalous Hall insulators and fractional Chern insulators 
when the superlattice potential modulation is produced by dielectric patterning or adjacent moir\'e materials.
\end{abstract}

\maketitle

\section{Introduction}

Moir\'e materials including twisted graphene \cite{bistritzer2011moire, cao2018correlated, cao2018unconventional, yankowitz2019tuning, serlin2020intrinsic, xie2021fractional, andrei2020graphene, liu2020tunable, park2021tunable, lu2023fractional} and transition metal dichalcogenide (TMD) \cite{wu2018hubbard, wu2019topological, devakul2021magic, tang2020simulation, regan2020mott, xu2020correlated, ghiotto2021quantum, li2021quantum, tao2022valley, mak2022semiconductor, zeng2023thermodynamic, cai2023signatures, park2023observation, xu2023observation} heterostructures have attracted a tremendous amount of research interest. Motivated by the rich phenomenology observed in moir\'e materials, alternate flat-band platforms \cite{kennes2021moire, ghorashi2023topological, ghorashi2023multilayer, crepel2023chiral, gao2023untwisting,wan2023topological} have been proposed. Topological flat-band systems are especially interesting as the interplay between correlation and topology leads to quantum anomalous Hall \cite{serlin2020intrinsic, li2021quantum} and fractional Chern insulators \cite{xie2021fractional, zeng2023thermodynamic, cai2023signatures, park2023observation, lu2023fractional} at integer and fractional fillings, respectively. Flat bands with non-zero Chern numbers
are reminiscent of Landau levels in quantum Hall systems \cite{tarnopolsky2019origin, liu2019pseudo, bultinck2020mechanism, ledwith2021strong,wang2021exact}, and could underlie many of
the exotic phenomena observed in moir\'e materials, including superconductivity \cite{khalaf2021charged, chatterjee2022skyrmion, torma2022superconductivity}.

It was shown in a recent study \cite{ghorashi2023topological} that Bernal stacked bilayer graphene modulated by a superlattice potential (SL-BLG) provides a versatile platform for the study of flat-band physics. Band structure calculations show that SL-BLG hosts topologically nontrivial flat bands under weak superlattice potential modulation, and a stack of trivial flat bands under strong modulation. The flat bands in different parameter regimes are reminiscent of those in graphene and TMD moir\'e materials, but the high tunability makes SL-BLG a unique platform for the study of correlated and topological phases.

In this paper we study the origin of topological flat bands in the weak modulation limit, and the nature of correlated insulating states at integer filling of the topological flat bands. We find that when the superlattice potential is weak compared to displacement field, the lowest miniband above charge neutrality is always topologically nontrivial when the superlattice potential minima form a honeycomb lattice. At filling factor $\nu \equiv NA_0/A= 1$ of the topological flat bands, the system is a valley-polarized quantum anomalous Hall insulator over a large parameter range. (Here $N$ is the number of 
electrons, $A$ is the area of the system, and $A_0$ is the area of one unit cell.) 

This paper is organized as follows. In Sec.~\ref{sec:band_topology}, we introduce our model system and show how its band structure and topology is influenced by the strength of displacement field and the shape of superlattice potential. In Sec.~\ref{sec:corr_state}, we study interaction effects at filling $\nu=1$ of the topological flat bands and present mean-field phase diagrams. Finally, we conclude our work in Sec.~\ref{sec:discussion} with a discussion of the prospects for experimental realization of quantum anomalous Hall and fractional Chern insulators within the SL-BLG platform.

\section{Band Topology} \label{sec:band_topology}

\subsection{Model}

\begin{figure}
    \centering
    \includegraphics[width=\linewidth]{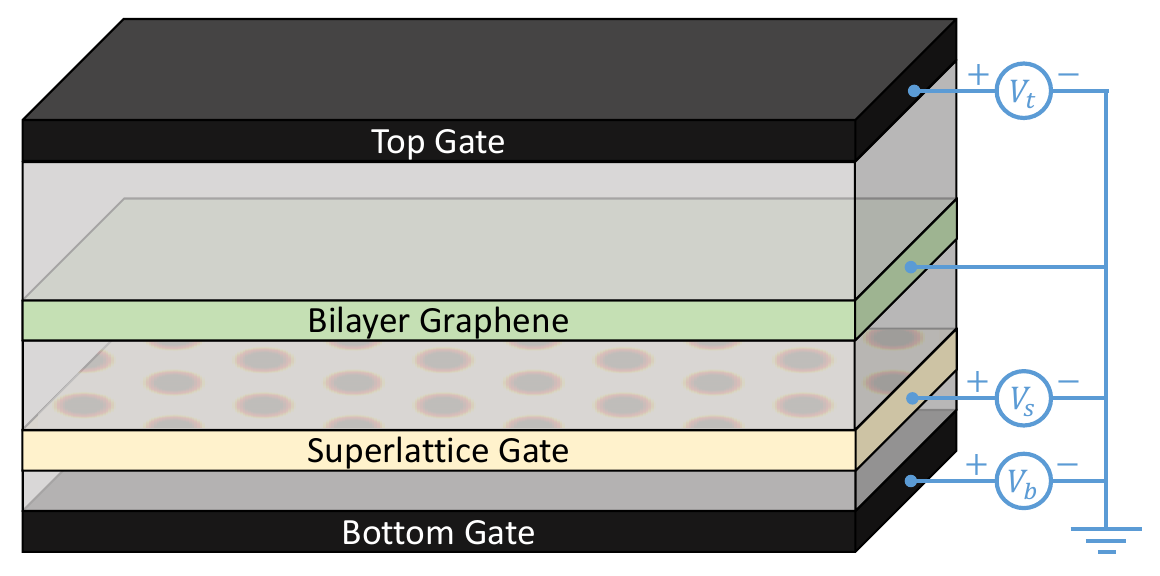}
    \caption{Schematic illustration of the SL-BLG device structure. The top and bottom gate voltages $V_t, V_b$ control the net charge density on bilayer graphene and the displacement field strength, while the superlattice gate voltage $V_s$ controls the superlattice potential strength. The gray regions represent dielectric layers that separate the active layers from the gates.}
    \label{fig:device}
\end{figure}

We consider Bernal-stacked bilayer graphene with a superlattice modulation potential (SL-BLG). For concreteness, we consider the experimental setup shown in Fig.~\ref{fig:device} in which the superlattice potential is produced by a periodic pattern on the superlattice gate. The gate-patterning technique 
depicted in Fig.~\ref{fig:device} has been demonstrated in recent experiments \cite{forsythe2018band, li2021anisotropic, barcons2022engineering, sun2023signature} and in principle allows superlattice potentials of arbitrary shapes and periodicities. By tuning the top and bottom gate voltages $V_t,V_b$ and the superlattice gate voltage $V_s$, the net charge density, displacement field, and superlattice potential strength can be independently controlled. We note that another efficient method to impose a superlattice potential is by including a twisted hexagonal boron nitride (hBN) bilayer,
which exhibits a a spatially periodic pattern of out-of-plane electric polarizations \cite{yasuda2021stacking, vizner2021interfacial, wang2022interfacial, kim2023electrostatic}, in 
the van der Waals layer stack.  A doped TMD moir\'e bilayer with localized charge carriers \cite{wang2022engineering} is yet another possibility, 
but for a general theoretical discussion we focus first on the highly tunable device structure in Fig.~\ref{fig:device} and defer the discussion of hBN and TMD bilayers to Sec.~\ref{sec:discussion}.

The system is described by the Hamiltonian
\begin{equation} \label{eq:Ham}
H = H_{\rm BLG} + H_{\rm SL} + H_{\rm int}.
\end{equation}
Here $H_{\rm BLG}$ is the Hamiltonian for bilayer graphene (BLG) with a displacement field:
\begin{equation}
\begin{split}
H_{\rm BLG} = \sum_{\tau s\bm k} \Psi_{\tau s\bm k}^{\dagger} &\Big[ \hbar v (\tau k_x \sigma_1 + k_y \sigma_2) \\
&+ \frac t2 (\rho_1 \sigma_1 - \rho_2 \sigma_2) + V_0 \rho_3 \Big] \Psi_{\tau s\bm k},
\end{split}
\end{equation}
where $\tau=\pm$ and $s=\uparrow,\downarrow$ are the valley and spin labels, $\Psi$ is a four-component spinor with layer and sublattice components, and  
$\rho$ and $\sigma$ are respectively layer and sublattice Pauli matrices, $v=\SI{e6}{m/s}$ is the Dirac velocity, $t=\SI{0.4}{eV}$ is the interlayer tunneling amplitude, and $V_0$ is the displacement field strength. $H_{\rm SL}$ describes the superlattice potential modulation:
\begin{equation}
H_{\rm SL} = \sum_{\tau s} \int d\bm r \, \Psi_{\tau s}^{\dagger}(\bm r) V(\bm r) \Psi_{\tau s}(\bm r).
\end{equation}
For most of this paper we focus on triangular superlattice potentials. To lowest order in the Fourier expansion,
\begin{equation}
V(\bm r) = \left( \frac{1+\alpha}{2}\rho_0 + \frac{1-\alpha}{2}\rho_3 \right) \left( \sum_{n=0}^2 V_{\rm SL}\, e^{i\bm g_n\cdot\bm r} + {\rm c.c.} \right),
\end{equation}
where in general $V_{\rm SL} = |V_{\rm SL}| e^{i\phi}$ is a complex parameter, and $\bm g_n = (4\pi/\sqrt{3}L)(\cos(2n\pi/3),\sin(2n\pi/3))$ with $n=0,1,2$ are reciprocal lattice vectors of the triangular superlattice with lattice constant $L$. The phase $\phi$ determines the ratios of the superlattice potential values at three high-symmetry points within a unit cell; they are respectively proportional to $\cos\phi$, $\cos(\phi+2\pi/3)$, and $\cos(\phi-2\pi/3)$. The lower two become degenerate at $\phi=2n\pi/3$ and at this point the potential minima form a honeycomb lattice. Similarly the potential maxima form a honeycomb lattice when $\phi=(2n+1)\pi/3$. $\alpha\in(0,1)$ is the ratio of effective superlattice potential strengths felt by the top and bottom graphene layers. In this paper we choose $\alpha=0.9$. Note that the value of $\alpha$ here is its {\em bare} value that only accounts for the geometric origin of layer asymmetry, {\it i.e.}, the fact that the bottom layer is closer to the superlattice gate than the top layer. Electrostatic screening leads to a much reduced effective $\alpha$ \cite{rokni2017layer}, but this effect is taken into account self-consistently by our Hartree-Fock calculations. To avoid double-counting, we start with a large bare $\alpha$ in the single-particle Hamiltonian. The precise value of $\alpha$ does not qualitatively affect our results. 

The last term in Eq.~\eqref{eq:Ham} is the Coulomb interaction
\begin{equation}
H_{\rm int} = \frac{1}{2A} \sum_{ll'\bm q} V_{ll'}(q) :n_{l,\bm q} n_{l',-\bm q}:,
\end{equation}
where $n_{l,\bm q} = \sum_{\sigma\tau s\bm k} a_{l\sigma\tau s,\bm k+\bm q}^{\dagger} a_{l\sigma\tau s,\bm k}$ is the density operator in layer $l$, $V_{ll'}(q) = 2\pi e^2/\epsilon q$ for $l=l'$ and $V_{ll'}(q) = (2\pi e^2/\epsilon q)e^{-qd}$ for $l\ne l'$ where $\epsilon$ is the dielectric constant and $d$ is the distance between the two graphene layers. The colons represent normal ordering of creation and annihilation operators. In our calculations we take $\epsilon=10$ and $d=\SI{3.5}{\angstrom}$. Intralayer and interlayer interactions must be distinguished in order to account for screening effects properly.

\subsection{Band structure and topology}

\begin{figure*}
    \includegraphics[width=0.29\linewidth,valign=t]{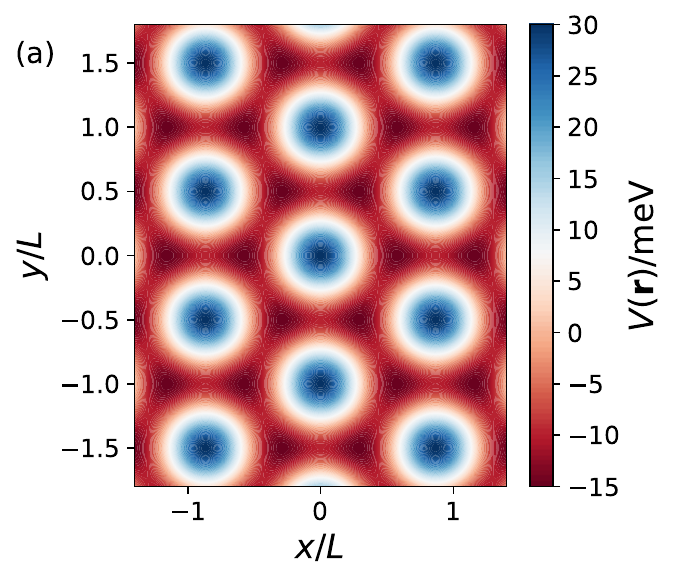}
    \includegraphics[width=0.32\linewidth,valign=t]{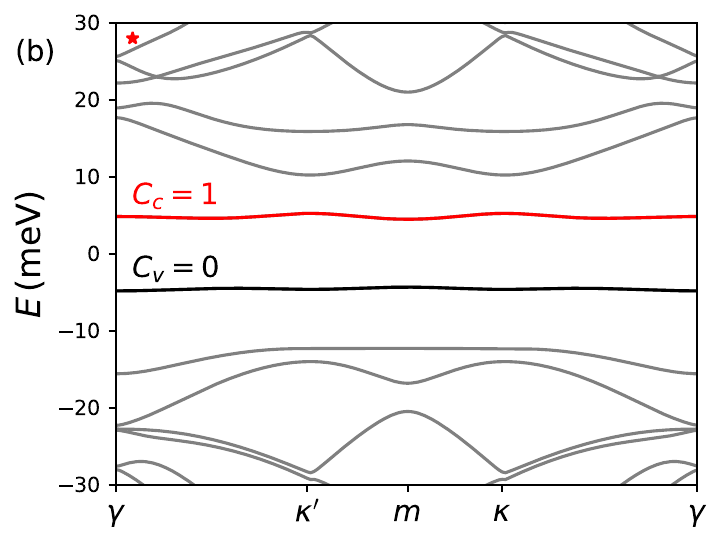}
    \includegraphics[width=0.32\linewidth,valign=t]{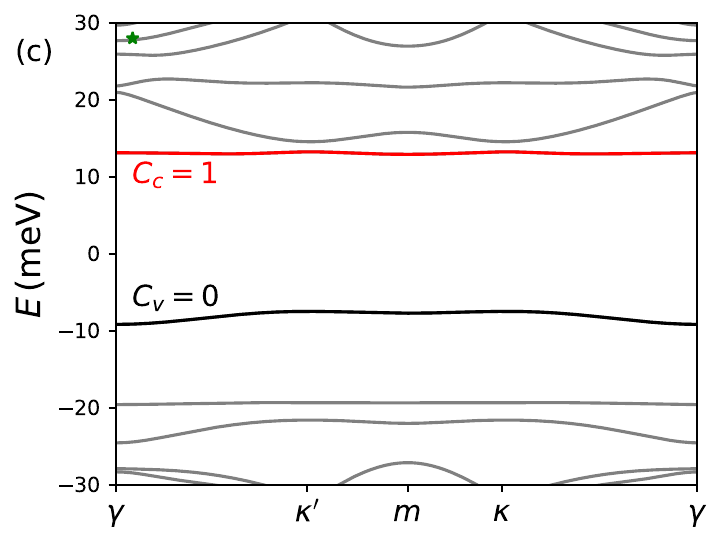}
    \includegraphics[width=0.29\linewidth,valign=t]{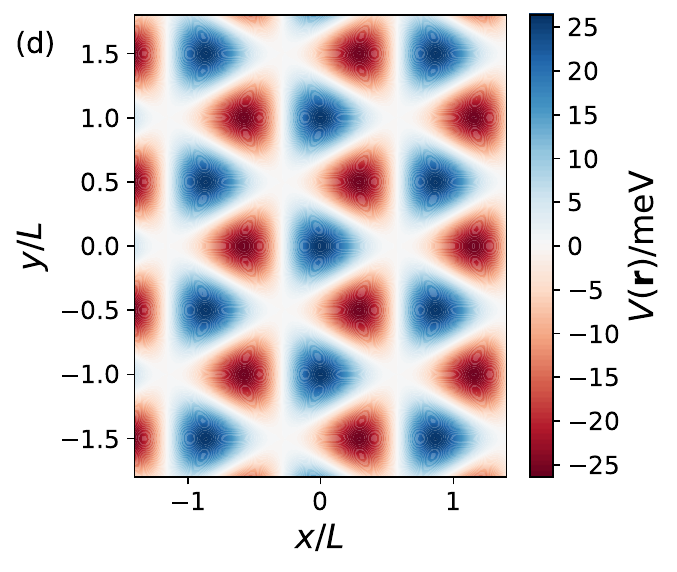}
    \includegraphics[width=0.32\linewidth,valign=t]{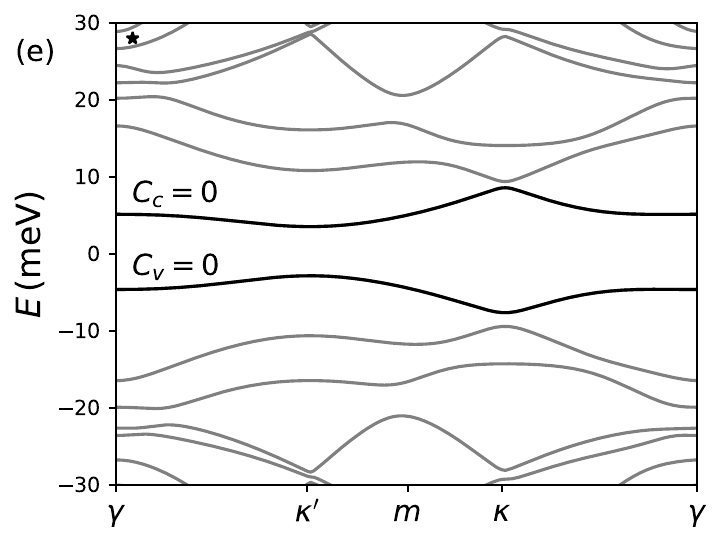}
    \includegraphics[width=0.34\linewidth,valign=t]{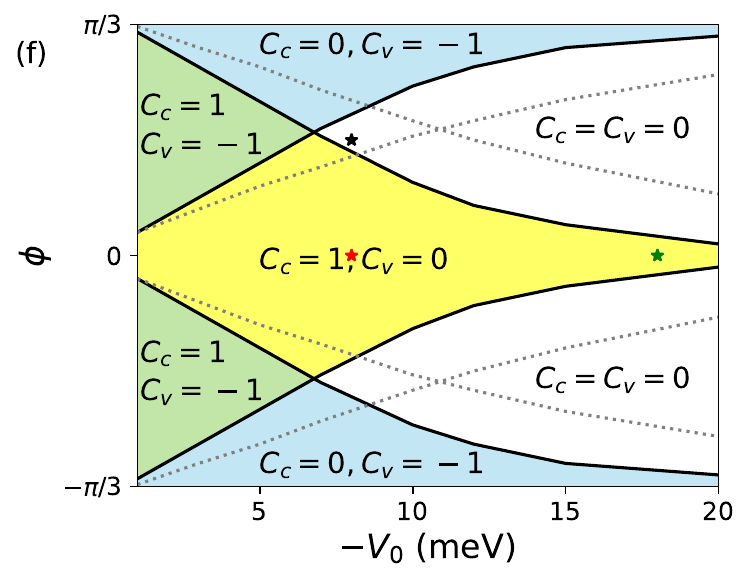}
    \caption{Band structure and topology of SL-BLG. (a) Superlattice potential with $|V_{\rm SL}| = \SI{5}{meV}$ and $\phi=0$. The potential minima form a honeycomb lattice and the potential maxima form a triangular lattice. (b,c) Band structure in valley $\tau=-1$ with displacement field (b) $V_0 = \SI{-8}{meV}$ (c) $V_0 = \SI{-18}{meV}$, with the superlattice potential in (a) and lattice constant $L=\SI{50}{nm}$. (d,e) Same plots as in (a,b) but with $\phi=\pi/6$. (f) Topological phase diagram in $V_0$-$\phi$ parameter space with fixed $|V_{\rm SL}| = \SI{5}{meV}$. $C_c$ and $C_v$ are the valley-projected Chern numbers of the first conduction and valence minibands respectively. The colored regions separated by solid lines represent different topological phases for $L=\SI{50}{nm}$, while the dotted lines are phase boundaries for $L=\SI{30}{nm}$. The red, green, and black stars represent the parameter choices in (b), (c), and (e) respectively.}
    \label{fig:bands}
\end{figure*}

The electronic band structure of SL-BLG has been studied in detail in Ref.~\cite{ghorashi2023topological}. Here, we briefly review some results and provide new insight 
into the nature of topological flat bands.

In the absence of a superlattice potential, a displacement field opens a $2|V_0|$ gap. 
The result is a valley Hall insulator with nonzero Berry curvature concentrated near the conduction band bottom and valence band top in each valley. 
With a weak superlattice potential ($|V_{\rm SL}|\ll |V_0|$), the conduction band electrons are localized near the potential minima and the valence band holes near the potential maxima. The potential minima form a triangular lattice for generic phase angles $\phi = \arg V_{\rm SL}$, but two local minima become degenerate at $\phi=2n\pi/3$ and together form a honeycomb lattice (Fig.~\ref{fig:bands}(a)). Similarly, the potential maxima form a honeycomb lattice near $\phi=(2n+1)\pi/3$ and a triangular lattice otherwise. We show below that lattice geometry is crucial for the topological properties of low-energy minibands; topological bands are favored by honeycomb lattices while triangular lattices are likely to host trivial bands.

Below we focus on the first miniband above charge neutrality. Physically this band represents the lowest bound state of conduction band electrons localized near the superlattice potential minima. Because the band dispersion of BLG is flat near the band extrema, the bandwidth of the first conduction miniband is small even with relatively weak potential modulation. To understand the topology of the band, we assume a weak superlattice potential and perform degenerate state perturbation theory near the $\kappa$-point [$\bm{\kappa} = (2\pi/3L) (\sqrt{3}, 1)$] of the mini Brillouin zone (mBZ). Folding the conduction band into the first mBZ, the lowest energy level at $\kappa$ is three-fold degenerate. The superlattice potential couples these three states and lifts the degeneracy. Projecting the superlattice potential onto the $3\times 3$ low-energy subspace, the Hamiltonian matrix reads
\begin{equation}
H_{\kappa} = \begin{pmatrix}
E_{\kappa} & V_{\kappa} & V_{\kappa}^* \\
V_{\kappa}^* & E_{\kappa} & V_{\kappa} \\
V_{\kappa} & V_{\kappa}^* & E_{\kappa}
\end{pmatrix},
\end{equation}
where $E_{\kappa}$ is the conduction band energy at $\kappa$ and $V_{\kappa} = \braket{\bm\kappa | V | \bm\kappa - \bm g_0}$ is the superlattice potential projected onto the low-energy subspace. If the basis states were trivial plane waves, $V_{\kappa} = V_{\rm SL}$ would simply be the Fourier coefficient of the superlattice potential. However, the BLG basis states have nontrivial spinor structures which 
produce an extra phase factor $e^{i\phi'}$, {\it i.e.}, $\arg V_{\kappa} = \phi+\phi'$. 
A similar analysis applies for the bands at $\kappa'=-\kappa$, but here the projected matrix element has the phase angle $\arg V_{\kappa'} = \phi-\phi'$.

The matrix $H_{\kappa}$ has three distinct eigenvalues in general, but the lower two eigenvalues become degenerate when $\arg V_{\kappa} = 2n\pi/3$. At this point the lower two bands form a Dirac cone near $\kappa$. By a similar argument the gap closes at $\kappa'$ when $\arg V_{\kappa'} = 2n\pi/3$. 
For a trivial band without any internal spinor structure, $\phi'=0$, the gap closes simultaneously at $\kappa$ and $\kappa'$ and the band topology does not change. In this case the lowest band is topologically trivial for any $\phi\ne 2n\pi/3$. When $\phi'\ne 0$, the gap closes at $\phi=2n\pi/3-\phi'$ at $\kappa$ and $\phi=2n\pi/3+\phi'$ at $\kappa'$, opening up an interval $\phi\in (2n\pi/3-\phi', 2n\pi/3+\phi')$ within which the lowest band is topologically nontrivial. The point $\phi=2n\pi/3$ at which the potential minima form a honeycomb lattice is at the middle of the topological region and is the optimal parameter choice for a topological flat band. This explains why in Ref.~\cite{ghorashi2023topological} (see also Fig.~\ref{fig:bands}(a-c)), where $\phi=0$ is implicitly assumed, the first miniband above charge neutrality is topologically nontrivial, while the band below, which originates from valence band holes localized at the potential maxima that form a triangular lattice, is topologically trivial.


The phase $\phi'$ originates from the nontrivial geometry of Bloch wavefunctions and determines the width of the topological region. Roughly speaking $\phi'$ is proportional to the Berry flux of the conduction band within the first mBZ. For small $\phi'$, gap closings at $\kappa$ and $\kappa'$ occur almost simultaneously as $\phi$ varies. While the lowest miniband is topologically nontrivial between these two gap closings, it is separated from the second lowest band only by a very small gap. For BLG, this is the case when the displacement field $V_0$ is large such that the Bloch wavefunctions near the conduction band bottom have little winding (see Fig.~\ref{fig:bands}(c)). In order to obtain an isolated topological flat band, $V_0$ cannot be too large.  An approximate criterion,
obtained by requiring a significant amount of Berry flux within the first mBZ, is that 
$V_0 \lesssim (\hbar^2 v^2/t)(2\pi/L)^2$. A smaller lattice constant $L$ implies a larger mBZ which contains a larger amount of Berry flux and therefore a larger $\phi'$ and a wider topological region.

Fig.~\ref{fig:bands}(f) shows the phase diagram in $V_0$-$\phi$ parameter space with fixed $|V_{\rm SL}| = \SI{5}{meV}$. In agreement with the above discussion, the topological region is centered at $\phi=0$ for the first conduction miniband and $\phi=\pi/3$ for the first valence miniband, and the width of the topological region shrinks rapidly with increasing $|V_0|$. The topological region becomes wider at smaller lattice constant $L$, as is clear from the comparison between the $L=\SI{50}{nm}$ (solid lines) and $L=\SI{30}{nm}$ (dotted lines) phase boundaries.

As the superlattice potential strength $|V_{\rm SL}|$ increases, the conduction minibands move down and the valence minibands move up. The conduction and valence minibands start to mix when $|V_{\rm SL}|\sim V_0$ and above this point the band structure becomes complicated and sensitive to parameters. In the strong modulation limit the system becomes a quantum dot array with electrons and holes confined in neighboring sites. In this paper, we focus on the regime of relatively weak superlattice potential in which the band structure and topology can be understood by the arguments above, and leave the study of the strong modulation limit to future work.

\section{Correlated states in SL-BLG} \label{sec:corr_state}

In this section we study correlated states in SL-BLG at partial filling of the minibands, with emphasis on the topology of correlated insulating states. Since topological bands are favored by honeycomb lattices, we choose $\phi=0$ and focus on the first conduction miniband. The effects of varying $\phi$ will also be discussed.

\subsection{Mean-field theory}

We study interaction effects by Hartree-Fock mean-field theory. Because the first conduction miniband is not flat and isolated for generic parameters $V_0$ and $V_{\rm SL}$, we do not project interactions onto the low-energy minibands, but instead perform mean-field calculations in the plane-wave basis. The mean-field Hamiltonian consists of single-particle terms and Hartree-Fock self-energies:
\begin{equation}
H_{\rm MF} = H_{\rm BLG} + H_{\rm SL} + \Sigma_H + \Sigma_F.
\end{equation}
The Hartree term is physically an electrostatic potential:
\begin{equation} \label{eq:Hartree}
\Sigma_H = \frac 1A \sum_{l\sigma\tau s} \sum_{\bm{gk}} \left[ \sum_{l'} V_{ll'}(\bm g) n_{l'\bm{g}} \right] a_{l\sigma\tau s, \bm{k}+\bm{g}}^{\dagger} a_{l\sigma\tau s,\bm{k}},
\end{equation}
where $n_{l\bm{g}} = \sum_{\sigma\tau s \bm{k}} \langle a_{l\sigma\tau s,\bm{k}-\bm{g}}^{\dagger} a_{l\sigma\tau s,\bm{k}} \rangle$ is the Fourier transform of electron density in layer $l$. 
The Hartree potential must be regularized because of the negative energy sea; we account for its effects by defining $\langle\dots\rangle$ in the self-consistent field equations
as the expectation value in the mean-field ground state subtracted by that in charge-neutral BLG in the absence of external fields. The Fock term is
\begin{equation} \label{eq:Fock}
\begin{split}
\Sigma_F = &-\frac 1A \sum_{l'\sigma'\tau' s'} \sum_{l\sigma\tau s} \sum_{\bm{kk'g}} V_{ll'}(\bm{k'} - \bm{k} - \bm{g}) \\
&\times \langle a_{l\sigma\tau s,\bm{k'}-\bm{g}}^{\dagger} a_{l'\sigma'\tau's', \bm{k'}} \rangle a_{l'\sigma'\tau's', \bm{k}+\bm{g}}^{\dagger} a_{l\sigma\tau s, \bm{k}}.
\end{split}
\end{equation}
In Eqs.~\eqref{eq:Hartree}-\eqref{eq:Fock} we have assumed that the 
translational symmetry of the superlattice is preserved. The mean-field ground state is obtained by solving the mean-field equations self-consistently.

\subsection{Phase diagrams}

\begin{figure*}
    \centering
    \includegraphics[width=0.32\linewidth]{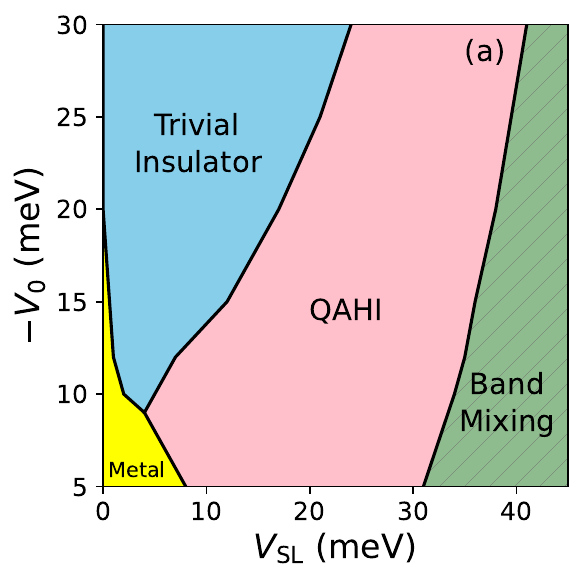}
    \includegraphics[width=0.32\linewidth]{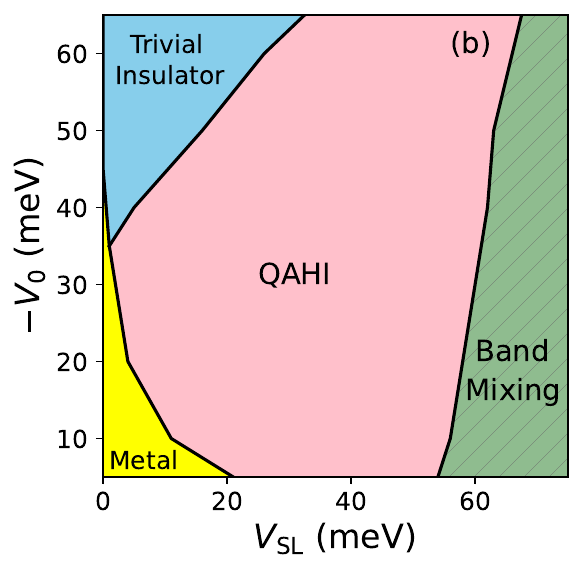}
    \includegraphics[width=0.33\linewidth]{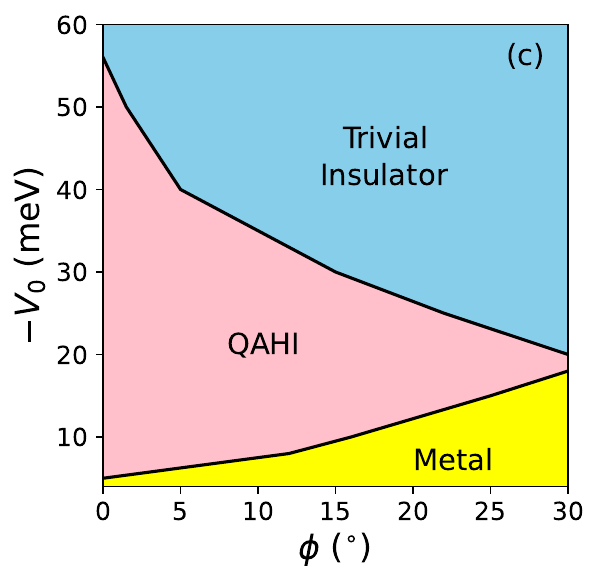}
    \caption{Mean-field phase diagrams at filling factor $\nu=1$. (a) and (b) are phase diagrams in displacement field $V_0$ and superlattice potential strength $|V_{\rm SL}|$ plane, with fixed $\phi=0$ but different lattice constants: (a) $L=\SI{50}{nm}$; (b) $L=\SI{30}{nm}$. The `band mixing' (hatched green) regions identify large-$|V_{\rm SL}|$ semimetal states in which conduction and valence minibands overlap and mix, leading to a complicated series of phase transitions between insulating and metallic states that we do not study in detail in this work. (c) is the phase diagram in $V_0$ and $\phi$ at fixed $|V_{\rm SL}| = \SI{20}{meV}$ and $L = \SI{30}{nm}$.}
    \label{fig:phase_diagrams}
\end{figure*}

The single-particle band structure has four-fold spin-valley degeneracy. 
Therefore at miniband filling $\nu=1$, the first conduction miniband of each flavor is quarter-filled, resulting in a metallic state, when interactions are absent.
This is indeed the case when the superlattice potential is weak so that the minibands remain dispersive. As the first miniband flattens with increasing $|V_{\rm SL}|$, it is energetically favorable to form a correlated insulating state by spontaneously breaking the spin-valley degeneracy. The threshold $|V_{\rm SL}|$ for correlated insulating states decreases with $|V_0|$ because increasing $|V_0|$ flattens the BLG bands. Interestingly, at large $|V_0|$ the correlated insulating state persists to the limit of vanishing $|V_{\rm SL}|$, pointing to the possibility of Wigner crystallization in BLG at large displacement field and low doping density \cite{silvestrov2017wigner, joy2023wigner}.

In our mean-field calculations we find that spin-valley-polarized states are energetically favored over intervalley-coherent states. If the only relevant band is the first conduction miniband and the 
shape parameter of the superlattice potential is near $\phi=0$, spin and valley polarized insulating states are quantum anomalous Hall insulators (QAHI) with Chern number $|C|=1$. This is the case when the first conduction miniband is isolated from all other bands by a gap larger than the interaction energy scale. According to the discussion in the last section, at $\phi=0$ the first conduction miniband is flat and isolated only at intermediate values of $|V_0|$ and $|V_{\rm SL}|$. At large $|V_0|$ the gap between the first and second minibands is small, and the Berry curvature is strongly localized near the $\kappa$ and $\kappa'$ points. At large $|V_{\rm SL}|$, on the other hand, mixing between the conduction and valence minibands becomes important, and the first conduction miniband is generally not flat or isolated.

The above picture is confirmed by the mean-field phase diagrams in Fig.~\ref{fig:phase_diagrams}. We find that the ground state at filling $\nu=1$ is a QAHI over a large region in voltage parameter space. At small $|V_0|$ the system is metallic at small $|V_{\rm SL}|$ and becomes a QAHI above a threshold $|V_{\rm SL}|$. The QAHI state persists until the conduction and valence minibands mix at large $|V_{\rm SL}|$, after which the system becomes either a trivial insulator or a metal. The phase diagram in the 
semimetal-like conduction/valence band-mixing regime (hatched green region in Fig.~\ref{fig:phase_diagrams}) at large $|V_{\rm SL}|$ is mostly topologically trivial.  Flavor symmetry breaking properties in this regime are highly sensitive to parameters, and in all likelihood not predicted correctly by Hartree-Fock theory.  
As $|V_0|$ increases, mixing between the first and second conduction minibands becomes increasingly important and leads to the emergence of a trivial insulating state between the metallic and QAHI states at large $|V_0|$. The trivial insulator region expands with increasing $|V_0|$, accompanied by the shrinking of the QAHI region.

Figs.~\ref{fig:phase_diagrams}(a) and (b) are phase diagrams at different lattice constants $L$. Comparison of the two shows that the QAHI region is significantly larger at smaller $L$. This is consistent with the single-particle phase diagram in Fig.~\ref{fig:bands}(e) in which the topological region widens at large $L$. Fig.~\ref{fig:phase_diagrams}(c) shows that the QAHI region quickly narrows
as $\phi$ increases from $0$ to $\pi/6$, in agreement with the insight from the last section that topological bands are favored by honeycomb lattices ($\phi=0$).

\section{Conclusion and discussion} \label{sec:discussion}

In this paper we studied the band structure and topology of SL-BLG as well as correlated insulating states that appear at partial filling of the first conduction miniband. We found that the ground state at filling $\nu=1$ is a spin-valley-polarized QAHI when the first conduction miniband is isolated, flat, and topologically nontrivial. According to the analysis in Sec.~\ref{sec:band_topology}, this is the case when: (i) the superlattice potential minima form a honeycomb lattice with two equivalent sublattices ({\it i.e.,} $\phi=0$); and (ii) the displacement field $V_0$ and superlattice potential strength $|V_{\rm SL}|$ are relatively weak. The large stable QAHI regions in our mean-field phase diagrams (Fig.~\ref{fig:phase_diagrams}) suggest that QAHIs should be routinely realized in honeycomb-superlattice potential modulated BLG with no need for fine-tuning.

Our gap-closing argument in Sec.~\ref{sec:band_topology} is not specific to BLG; generalization to other systems as well as other superlattice geometries is straightforward. We note that similar perturbative analysis was performed in Ref.~\cite{su2022massive} in the context of TMD moir\'e bilayers, where the moir\'e potential plays the role of superlattice potential in our work with $\phi$ tunable by electric field, and the TMD bands are modeled as massive Dirac fermions that carry intrinsic Berry curvature. Our argument shows that when weak honeycomb-lattice potential modulation is applied to any gapped system with nontrivial band geometry, the lowest miniband above the band gap is likely to be topologically nontrivial. Since the central point of our argument is built upon Berry phase effects that break the effective time-reversal symmetry of the single-valley Hamiltonian and gap out the Dirac cones at $\kappa,\kappa'$, our results are also applicable to kagome lattice potential modulations. Note, however, that gap closing at $\bm{\gamma}=(0,0)$ and $\bm{m} = (2\pi/\sqrt{3}L,0)$ can also change the topology of minibands. Neither case occurs in BLG, but both are possible for other systems such as the quantum spin Hall insulators described by the Bernevig-Hughes-Zhang model \cite{bernevig2006quantum}. For square-lattice potential modulated BLG, a weak modulation potential always opens up a trivial gap between the first and second conduction minibands. Topological phase transitions do occur when the modulation potential gets strong as shown in Ref.~\cite{ghorashi2023topological}, but this is outside the scope of our perturbation theory analysis, and the topological bands in this regime are not as robust and likely require more fine-tuning.
A more general analysis of topological band inversions in superlattice potential modulated two-dimensional systems as well as construction of lattice models is left for future work.

It was recently shown experimentally that superlattice modulation potentials can also be produced by adjacent moir\'e bilayers \cite{kim2023electrostatic, wang2022engineering}. Twisted hBN bilayers produce a triangular superlattice potential \cite{zhao2021universal} with $\phi=\pi/6$, which according to our phase diagrams in Fig.~\ref{fig:bands}(e) and Fig.~\ref{fig:phase_diagrams}(c) is unfavorable for topological flat bands \cite{brzezinska2023flat} and QAHIs. TMD bilayers, on the other hand, provide more tunability with choice of material combinations and doping densities. Twisted TMD homobilayers \cite{angeli2021gamma} provide a honeycomb-lattice moir\'e potential for $\Gamma$-valley holes, and when the honeycomb lattice is filled by doping, the modulation potential for adjacent layers is honeycomb-shaped. The gate-patterning technique provides higher tunability on the shape and strength of the superlattice potential, but at the current stage it is limited to relatively long superlattice periodicities ($\gtrsim\SI{30}{nm}$), and the fabrication process introduces significant disorder.

It was shown in Ref.~\cite{ghorashi2023topological} that the topological flat band of SL-BLG in the weak modulation regime has nearly ideal quantum geometry, and that the ground state at filling $\nu=1/3$ is a fractional Chern insulator (FCI). The robustness of the topological flat bands and related valley-polarized QAHIs shown in our work suggests that the validity of these results is probably not limited to a special parameter choice, but that similar results are expected over a wide parameter range. Since the formation of FCIs requires spontaneous valley polarization, we expect that observation of FCIs at fractional fillings is likely within the QAHI regions in Fig.~\ref{fig:phase_diagrams}. A detailed study of FCIs in SL-BLG and related superlattice modulated systems is left for future work.

\begin{acknowledgements}
Y.Z. thanks Aidan Reddy for helpful discussion. Y.Z. acknowledges support from Programmable Quantum Materials, an Energy Frontiers Research Center funded by the U.S. Department of Energy (DOE), Office of Science, Basic Energy Sciences (BES), under award DE-SC0019443. Work at Austin was supported by the Department of Energy under grant DE-SC0019481. T.M.R.W. acknowledges support from the SNSF (Postdoc.Mobility\ No.~203152). 
J.C. and S.A.A.G. acknowledge support from the Air Force Office of Scientific Research under Grant No. FA9550-20-1-0260. J.C. is partially supported by the Alfred P. Sloan Foundation through a Sloan Research Fellowship.
The Flatiron Institute is a division of the Simons Foundation.
    
\end{acknowledgements}

\bibliography{ref.bib}

\begin{thebibliography}{55}%
\makeatletter
\providecommand \@ifxundefined [1]{%
 \@ifx{#1\undefined}
}%
\providecommand \@ifnum [1]{%
 \ifnum #1\expandafter \@firstoftwo
 \else \expandafter \@secondoftwo
 \fi
}%
\providecommand \@ifx [1]{%
 \ifx #1\expandafter \@firstoftwo
 \else \expandafter \@secondoftwo
 \fi
}%
\providecommand \natexlab [1]{#1}%
\providecommand \enquote  [1]{``#1''}%
\providecommand \bibnamefont  [1]{#1}%
\providecommand \bibfnamefont [1]{#1}%
\providecommand \citenamefont [1]{#1}%
\providecommand \href@noop [0]{\@secondoftwo}%
\providecommand \href [0]{\begingroup \@sanitize@url \@href}%
\providecommand \@href[1]{\@@startlink{#1}\@@href}%
\providecommand \@@href[1]{\endgroup#1\@@endlink}%
\providecommand \@sanitize@url [0]{\catcode `\\12\catcode `\$12\catcode
  `\&12\catcode `\#12\catcode `\^12\catcode `\_12\catcode `\%12\relax}%
\providecommand \@@startlink[1]{}%
\providecommand \@@endlink[0]{}%
\providecommand \url  [0]{\begingroup\@sanitize@url \@url }%
\providecommand \@url [1]{\endgroup\@href {#1}{\urlprefix }}%
\providecommand \urlprefix  [0]{URL }%
\providecommand \Eprint [0]{\href }%
\providecommand \doibase [0]{https://doi.org/}%
\providecommand \selectlanguage [0]{\@gobble}%
\providecommand \bibinfo  [0]{\@secondoftwo}%
\providecommand \bibfield  [0]{\@secondoftwo}%
\providecommand \translation [1]{[#1]}%
\providecommand \BibitemOpen [0]{}%
\providecommand \bibitemStop [0]{}%
\providecommand \bibitemNoStop [0]{.\EOS\space}%
\providecommand \EOS [0]{\spacefactor3000\relax}%
\providecommand \BibitemShut  [1]{\csname bibitem#1\endcsname}%
\let\auto@bib@innerbib\@empty
\bibitem [{\citenamefont {Bistritzer}\ and\ \citenamefont
  {MacDonald}(2011)}]{bistritzer2011moire}%
  \BibitemOpen
  \bibfield  {author} {\bibinfo {author} {\bibfnamefont {R.}~\bibnamefont
  {Bistritzer}}\ and\ \bibinfo {author} {\bibfnamefont {A.~H.}\ \bibnamefont
  {MacDonald}},\ }\bibfield  {title} {\bibinfo {title} {Moir{\'e} bands in
  twisted double-layer graphene},\ }\href@noop {} {\bibfield  {journal}
  {\bibinfo  {journal} {Proceedings of the National Academy of Sciences}\
  }\textbf {\bibinfo {volume} {108}},\ \bibinfo {pages} {12233} (\bibinfo
  {year} {2011})}\BibitemShut {NoStop}%
\bibitem [{\citenamefont {Cao}\ \emph {et~al.}(2018{\natexlab{a}})\citenamefont
  {Cao}, \citenamefont {Fatemi}, \citenamefont {Demir}, \citenamefont {Fang},
  \citenamefont {Tomarken}, \citenamefont {Luo}, \citenamefont
  {Sanchez-Yamagishi}, \citenamefont {Watanabe}, \citenamefont {Taniguchi},
  \citenamefont {Kaxiras} \emph {et~al.}}]{cao2018correlated}%
  \BibitemOpen
  \bibfield  {author} {\bibinfo {author} {\bibfnamefont {Y.}~\bibnamefont
  {Cao}}, \bibinfo {author} {\bibfnamefont {V.}~\bibnamefont {Fatemi}},
  \bibinfo {author} {\bibfnamefont {A.}~\bibnamefont {Demir}}, \bibinfo
  {author} {\bibfnamefont {S.}~\bibnamefont {Fang}}, \bibinfo {author}
  {\bibfnamefont {S.~L.}\ \bibnamefont {Tomarken}}, \bibinfo {author}
  {\bibfnamefont {J.~Y.}\ \bibnamefont {Luo}}, \bibinfo {author} {\bibfnamefont
  {J.~D.}\ \bibnamefont {Sanchez-Yamagishi}}, \bibinfo {author} {\bibfnamefont
  {K.}~\bibnamefont {Watanabe}}, \bibinfo {author} {\bibfnamefont
  {T.}~\bibnamefont {Taniguchi}}, \bibinfo {author} {\bibfnamefont
  {E.}~\bibnamefont {Kaxiras}}, \emph {et~al.},\ }\bibfield  {title} {\bibinfo
  {title} {Correlated insulator behaviour at half-filling in magic-angle
  graphene superlattices},\ }\href@noop {} {\bibfield  {journal} {\bibinfo
  {journal} {Nature}\ }\textbf {\bibinfo {volume} {556}},\ \bibinfo {pages}
  {80} (\bibinfo {year} {2018}{\natexlab{a}})}\BibitemShut {NoStop}%
\bibitem [{\citenamefont {Cao}\ \emph {et~al.}(2018{\natexlab{b}})\citenamefont
  {Cao}, \citenamefont {Fatemi}, \citenamefont {Fang}, \citenamefont
  {Watanabe}, \citenamefont {Taniguchi}, \citenamefont {Kaxiras},\ and\
  \citenamefont {Jarillo-Herrero}}]{cao2018unconventional}%
  \BibitemOpen
  \bibfield  {author} {\bibinfo {author} {\bibfnamefont {Y.}~\bibnamefont
  {Cao}}, \bibinfo {author} {\bibfnamefont {V.}~\bibnamefont {Fatemi}},
  \bibinfo {author} {\bibfnamefont {S.}~\bibnamefont {Fang}}, \bibinfo {author}
  {\bibfnamefont {K.}~\bibnamefont {Watanabe}}, \bibinfo {author}
  {\bibfnamefont {T.}~\bibnamefont {Taniguchi}}, \bibinfo {author}
  {\bibfnamefont {E.}~\bibnamefont {Kaxiras}},\ and\ \bibinfo {author}
  {\bibfnamefont {P.}~\bibnamefont {Jarillo-Herrero}},\ }\bibfield  {title}
  {\bibinfo {title} {Unconventional superconductivity in magic-angle graphene
  superlattices},\ }\href@noop {} {\bibfield  {journal} {\bibinfo  {journal}
  {Nature}\ }\textbf {\bibinfo {volume} {556}},\ \bibinfo {pages} {43}
  (\bibinfo {year} {2018}{\natexlab{b}})}\BibitemShut {NoStop}%
\bibitem [{\citenamefont {Yankowitz}\ \emph {et~al.}(2019)\citenamefont
  {Yankowitz}, \citenamefont {Chen}, \citenamefont {Polshyn}, \citenamefont
  {Zhang}, \citenamefont {Watanabe}, \citenamefont {Taniguchi}, \citenamefont
  {Graf}, \citenamefont {Young},\ and\ \citenamefont
  {Dean}}]{yankowitz2019tuning}%
  \BibitemOpen
  \bibfield  {author} {\bibinfo {author} {\bibfnamefont {M.}~\bibnamefont
  {Yankowitz}}, \bibinfo {author} {\bibfnamefont {S.}~\bibnamefont {Chen}},
  \bibinfo {author} {\bibfnamefont {H.}~\bibnamefont {Polshyn}}, \bibinfo
  {author} {\bibfnamefont {Y.}~\bibnamefont {Zhang}}, \bibinfo {author}
  {\bibfnamefont {K.}~\bibnamefont {Watanabe}}, \bibinfo {author}
  {\bibfnamefont {T.}~\bibnamefont {Taniguchi}}, \bibinfo {author}
  {\bibfnamefont {D.}~\bibnamefont {Graf}}, \bibinfo {author} {\bibfnamefont
  {A.~F.}\ \bibnamefont {Young}},\ and\ \bibinfo {author} {\bibfnamefont
  {C.~R.}\ \bibnamefont {Dean}},\ }\bibfield  {title} {\bibinfo {title} {Tuning
  superconductivity in twisted bilayer graphene},\ }\href@noop {} {\bibfield
  {journal} {\bibinfo  {journal} {Science}\ }\textbf {\bibinfo {volume}
  {363}},\ \bibinfo {pages} {1059} (\bibinfo {year} {2019})}\BibitemShut
  {NoStop}%
\bibitem [{\citenamefont {Serlin}\ \emph {et~al.}(2020)\citenamefont {Serlin},
  \citenamefont {Tschirhart}, \citenamefont {Polshyn}, \citenamefont {Zhang},
  \citenamefont {Zhu}, \citenamefont {Watanabe}, \citenamefont {Taniguchi},
  \citenamefont {Balents},\ and\ \citenamefont {Young}}]{serlin2020intrinsic}%
  \BibitemOpen
  \bibfield  {author} {\bibinfo {author} {\bibfnamefont {M.}~\bibnamefont
  {Serlin}}, \bibinfo {author} {\bibfnamefont {C.}~\bibnamefont {Tschirhart}},
  \bibinfo {author} {\bibfnamefont {H.}~\bibnamefont {Polshyn}}, \bibinfo
  {author} {\bibfnamefont {Y.}~\bibnamefont {Zhang}}, \bibinfo {author}
  {\bibfnamefont {J.}~\bibnamefont {Zhu}}, \bibinfo {author} {\bibfnamefont
  {K.}~\bibnamefont {Watanabe}}, \bibinfo {author} {\bibfnamefont
  {T.}~\bibnamefont {Taniguchi}}, \bibinfo {author} {\bibfnamefont
  {L.}~\bibnamefont {Balents}},\ and\ \bibinfo {author} {\bibfnamefont
  {A.}~\bibnamefont {Young}},\ }\bibfield  {title} {\bibinfo {title} {Intrinsic
  quantized anomalous hall effect in a moir{\'e} heterostructure},\ }\href@noop
  {} {\bibfield  {journal} {\bibinfo  {journal} {Science}\ }\textbf {\bibinfo
  {volume} {367}},\ \bibinfo {pages} {900} (\bibinfo {year}
  {2020})}\BibitemShut {NoStop}%
\bibitem [{\citenamefont {Xie}\ \emph {et~al.}(2021)\citenamefont {Xie},
  \citenamefont {Pierce}, \citenamefont {Park}, \citenamefont {Parker},
  \citenamefont {Khalaf}, \citenamefont {Ledwith}, \citenamefont {Cao},
  \citenamefont {Lee}, \citenamefont {Chen}, \citenamefont {Forrester} \emph
  {et~al.}}]{xie2021fractional}%
  \BibitemOpen
  \bibfield  {author} {\bibinfo {author} {\bibfnamefont {Y.}~\bibnamefont
  {Xie}}, \bibinfo {author} {\bibfnamefont {A.~T.}\ \bibnamefont {Pierce}},
  \bibinfo {author} {\bibfnamefont {J.~M.}\ \bibnamefont {Park}}, \bibinfo
  {author} {\bibfnamefont {D.~E.}\ \bibnamefont {Parker}}, \bibinfo {author}
  {\bibfnamefont {E.}~\bibnamefont {Khalaf}}, \bibinfo {author} {\bibfnamefont
  {P.}~\bibnamefont {Ledwith}}, \bibinfo {author} {\bibfnamefont
  {Y.}~\bibnamefont {Cao}}, \bibinfo {author} {\bibfnamefont {S.~H.}\
  \bibnamefont {Lee}}, \bibinfo {author} {\bibfnamefont {S.}~\bibnamefont
  {Chen}}, \bibinfo {author} {\bibfnamefont {P.~R.}\ \bibnamefont {Forrester}},
  \emph {et~al.},\ }\bibfield  {title} {\bibinfo {title} {Fractional chern
  insulators in magic-angle twisted bilayer graphene},\ }\href@noop {}
  {\bibfield  {journal} {\bibinfo  {journal} {Nature}\ }\textbf {\bibinfo
  {volume} {600}},\ \bibinfo {pages} {439} (\bibinfo {year}
  {2021})}\BibitemShut {NoStop}%
\bibitem [{\citenamefont {Andrei}\ and\ \citenamefont
  {MacDonald}(2020)}]{andrei2020graphene}%
  \BibitemOpen
  \bibfield  {author} {\bibinfo {author} {\bibfnamefont {E.~Y.}\ \bibnamefont
  {Andrei}}\ and\ \bibinfo {author} {\bibfnamefont {A.~H.}\ \bibnamefont
  {MacDonald}},\ }\bibfield  {title} {\bibinfo {title} {Graphene bilayers with
  a twist},\ }\href@noop {} {\bibfield  {journal} {\bibinfo  {journal} {Nature
  materials}\ }\textbf {\bibinfo {volume} {19}},\ \bibinfo {pages} {1265}
  (\bibinfo {year} {2020})}\BibitemShut {NoStop}%
\bibitem [{\citenamefont {Liu}\ \emph {et~al.}(2020)\citenamefont {Liu},
  \citenamefont {Hao}, \citenamefont {Khalaf}, \citenamefont {Lee},
  \citenamefont {Ronen}, \citenamefont {Yoo}, \citenamefont {Haei~Najafabadi},
  \citenamefont {Watanabe}, \citenamefont {Taniguchi}, \citenamefont
  {Vishwanath} \emph {et~al.}}]{liu2020tunable}%
  \BibitemOpen
  \bibfield  {author} {\bibinfo {author} {\bibfnamefont {X.}~\bibnamefont
  {Liu}}, \bibinfo {author} {\bibfnamefont {Z.}~\bibnamefont {Hao}}, \bibinfo
  {author} {\bibfnamefont {E.}~\bibnamefont {Khalaf}}, \bibinfo {author}
  {\bibfnamefont {J.~Y.}\ \bibnamefont {Lee}}, \bibinfo {author} {\bibfnamefont
  {Y.}~\bibnamefont {Ronen}}, \bibinfo {author} {\bibfnamefont
  {H.}~\bibnamefont {Yoo}}, \bibinfo {author} {\bibfnamefont {D.}~\bibnamefont
  {Haei~Najafabadi}}, \bibinfo {author} {\bibfnamefont {K.}~\bibnamefont
  {Watanabe}}, \bibinfo {author} {\bibfnamefont {T.}~\bibnamefont {Taniguchi}},
  \bibinfo {author} {\bibfnamefont {A.}~\bibnamefont {Vishwanath}}, \emph
  {et~al.},\ }\bibfield  {title} {\bibinfo {title} {Tunable spin-polarized
  correlated states in twisted double bilayer graphene},\ }\href@noop {}
  {\bibfield  {journal} {\bibinfo  {journal} {Nature}\ }\textbf {\bibinfo
  {volume} {583}},\ \bibinfo {pages} {221} (\bibinfo {year}
  {2020})}\BibitemShut {NoStop}%
\bibitem [{\citenamefont {Park}\ \emph {et~al.}(2021)\citenamefont {Park},
  \citenamefont {Cao}, \citenamefont {Watanabe}, \citenamefont {Taniguchi},\
  and\ \citenamefont {Jarillo-Herrero}}]{park2021tunable}%
  \BibitemOpen
  \bibfield  {author} {\bibinfo {author} {\bibfnamefont {J.~M.}\ \bibnamefont
  {Park}}, \bibinfo {author} {\bibfnamefont {Y.}~\bibnamefont {Cao}}, \bibinfo
  {author} {\bibfnamefont {K.}~\bibnamefont {Watanabe}}, \bibinfo {author}
  {\bibfnamefont {T.}~\bibnamefont {Taniguchi}},\ and\ \bibinfo {author}
  {\bibfnamefont {P.}~\bibnamefont {Jarillo-Herrero}},\ }\bibfield  {title}
  {\bibinfo {title} {Tunable strongly coupled superconductivity in magic-angle
  twisted trilayer graphene},\ }\href@noop {} {\bibfield  {journal} {\bibinfo
  {journal} {Nature}\ }\textbf {\bibinfo {volume} {590}},\ \bibinfo {pages}
  {249} (\bibinfo {year} {2021})}\BibitemShut {NoStop}%
\bibitem [{\citenamefont {Lu}\ \emph {et~al.}(2023)\citenamefont {Lu},
  \citenamefont {Han}, \citenamefont {Yao}, \citenamefont {Reddy},
  \citenamefont {Yang}, \citenamefont {Seo}, \citenamefont {Watanabe},
  \citenamefont {Taniguchi}, \citenamefont {Fu},\ and\ \citenamefont
  {Ju}}]{lu2023fractional}%
  \BibitemOpen
  \bibfield  {author} {\bibinfo {author} {\bibfnamefont {Z.}~\bibnamefont
  {Lu}}, \bibinfo {author} {\bibfnamefont {T.}~\bibnamefont {Han}}, \bibinfo
  {author} {\bibfnamefont {Y.}~\bibnamefont {Yao}}, \bibinfo {author}
  {\bibfnamefont {A.~P.}\ \bibnamefont {Reddy}}, \bibinfo {author}
  {\bibfnamefont {J.}~\bibnamefont {Yang}}, \bibinfo {author} {\bibfnamefont
  {J.}~\bibnamefont {Seo}}, \bibinfo {author} {\bibfnamefont {K.}~\bibnamefont
  {Watanabe}}, \bibinfo {author} {\bibfnamefont {T.}~\bibnamefont {Taniguchi}},
  \bibinfo {author} {\bibfnamefont {L.}~\bibnamefont {Fu}},\ and\ \bibinfo
  {author} {\bibfnamefont {L.}~\bibnamefont {Ju}},\ }\bibfield  {title}
  {\bibinfo {title} {Fractional quantum anomalous hall effect in a graphene
  moire superlattice},\ }\href@noop {} {\bibfield  {journal} {\bibinfo
  {journal} {arXiv preprint arXiv:2309.17436}\ } (\bibinfo {year}
  {2023})}\BibitemShut {NoStop}%
\bibitem [{\citenamefont {Wu}\ \emph {et~al.}(2018)\citenamefont {Wu},
  \citenamefont {Lovorn}, \citenamefont {Tutuc},\ and\ \citenamefont
  {MacDonald}}]{wu2018hubbard}%
  \BibitemOpen
  \bibfield  {author} {\bibinfo {author} {\bibfnamefont {F.}~\bibnamefont
  {Wu}}, \bibinfo {author} {\bibfnamefont {T.}~\bibnamefont {Lovorn}}, \bibinfo
  {author} {\bibfnamefont {E.}~\bibnamefont {Tutuc}},\ and\ \bibinfo {author}
  {\bibfnamefont {A.~H.}\ \bibnamefont {MacDonald}},\ }\bibfield  {title}
  {\bibinfo {title} {Hubbard model physics in transition metal dichalcogenide
  moir\'e bands},\ }\href {https://doi.org/10.1103/PhysRevLett.121.026402}
  {\bibfield  {journal} {\bibinfo  {journal} {Phys. Rev. Lett.}\ }\textbf
  {\bibinfo {volume} {121}},\ \bibinfo {pages} {026402} (\bibinfo {year}
  {2018})}\BibitemShut {NoStop}%
\bibitem [{\citenamefont {Wu}\ \emph {et~al.}(2019)\citenamefont {Wu},
  \citenamefont {Lovorn}, \citenamefont {Tutuc}, \citenamefont {Martin},\ and\
  \citenamefont {MacDonald}}]{wu2019topological}%
  \BibitemOpen
  \bibfield  {author} {\bibinfo {author} {\bibfnamefont {F.}~\bibnamefont
  {Wu}}, \bibinfo {author} {\bibfnamefont {T.}~\bibnamefont {Lovorn}}, \bibinfo
  {author} {\bibfnamefont {E.}~\bibnamefont {Tutuc}}, \bibinfo {author}
  {\bibfnamefont {I.}~\bibnamefont {Martin}},\ and\ \bibinfo {author}
  {\bibfnamefont {A.~H.}\ \bibnamefont {MacDonald}},\ }\bibfield  {title}
  {\bibinfo {title} {Topological insulators in twisted transition metal
  dichalcogenide homobilayers},\ }\href
  {https://doi.org/10.1103/PhysRevLett.122.086402} {\bibfield  {journal}
  {\bibinfo  {journal} {Phys. Rev. Lett.}\ }\textbf {\bibinfo {volume} {122}},\
  \bibinfo {pages} {086402} (\bibinfo {year} {2019})}\BibitemShut {NoStop}%
\bibitem [{\citenamefont {Devakul}\ \emph {et~al.}(2021)\citenamefont
  {Devakul}, \citenamefont {Cr{\'e}pel}, \citenamefont {Zhang},\ and\
  \citenamefont {Fu}}]{devakul2021magic}%
  \BibitemOpen
  \bibfield  {author} {\bibinfo {author} {\bibfnamefont {T.}~\bibnamefont
  {Devakul}}, \bibinfo {author} {\bibfnamefont {V.}~\bibnamefont {Cr{\'e}pel}},
  \bibinfo {author} {\bibfnamefont {Y.}~\bibnamefont {Zhang}},\ and\ \bibinfo
  {author} {\bibfnamefont {L.}~\bibnamefont {Fu}},\ }\bibfield  {title}
  {\bibinfo {title} {Magic in twisted transition metal dichalcogenide
  bilayers},\ }\href@noop {} {\bibfield  {journal} {\bibinfo  {journal} {Nature
  communications}\ }\textbf {\bibinfo {volume} {12}},\ \bibinfo {pages} {6730}
  (\bibinfo {year} {2021})}\BibitemShut {NoStop}%
\bibitem [{\citenamefont {Tang}\ \emph {et~al.}(2020)\citenamefont {Tang},
  \citenamefont {Li}, \citenamefont {Li}, \citenamefont {Xu}, \citenamefont
  {Liu}, \citenamefont {Barmak}, \citenamefont {Watanabe}, \citenamefont
  {Taniguchi}, \citenamefont {MacDonald}, \citenamefont {Shan} \emph
  {et~al.}}]{tang2020simulation}%
  \BibitemOpen
  \bibfield  {author} {\bibinfo {author} {\bibfnamefont {Y.}~\bibnamefont
  {Tang}}, \bibinfo {author} {\bibfnamefont {L.}~\bibnamefont {Li}}, \bibinfo
  {author} {\bibfnamefont {T.}~\bibnamefont {Li}}, \bibinfo {author}
  {\bibfnamefont {Y.}~\bibnamefont {Xu}}, \bibinfo {author} {\bibfnamefont
  {S.}~\bibnamefont {Liu}}, \bibinfo {author} {\bibfnamefont {K.}~\bibnamefont
  {Barmak}}, \bibinfo {author} {\bibfnamefont {K.}~\bibnamefont {Watanabe}},
  \bibinfo {author} {\bibfnamefont {T.}~\bibnamefont {Taniguchi}}, \bibinfo
  {author} {\bibfnamefont {A.~H.}\ \bibnamefont {MacDonald}}, \bibinfo {author}
  {\bibfnamefont {J.}~\bibnamefont {Shan}}, \emph {et~al.},\ }\bibfield
  {title} {\bibinfo {title} {Simulation of hubbard model physics in wse2/ws2
  moir{\'e} superlattices},\ }\href@noop {} {\bibfield  {journal} {\bibinfo
  {journal} {Nature}\ }\textbf {\bibinfo {volume} {579}},\ \bibinfo {pages}
  {353} (\bibinfo {year} {2020})}\BibitemShut {NoStop}%
\bibitem [{\citenamefont {Regan}\ \emph {et~al.}(2020)\citenamefont {Regan},
  \citenamefont {Wang}, \citenamefont {Jin}, \citenamefont {Bakti~Utama},
  \citenamefont {Gao}, \citenamefont {Wei}, \citenamefont {Zhao}, \citenamefont
  {Zhao}, \citenamefont {Zhang}, \citenamefont {Yumigeta} \emph
  {et~al.}}]{regan2020mott}%
  \BibitemOpen
  \bibfield  {author} {\bibinfo {author} {\bibfnamefont {E.~C.}\ \bibnamefont
  {Regan}}, \bibinfo {author} {\bibfnamefont {D.}~\bibnamefont {Wang}},
  \bibinfo {author} {\bibfnamefont {C.}~\bibnamefont {Jin}}, \bibinfo {author}
  {\bibfnamefont {M.~I.}\ \bibnamefont {Bakti~Utama}}, \bibinfo {author}
  {\bibfnamefont {B.}~\bibnamefont {Gao}}, \bibinfo {author} {\bibfnamefont
  {X.}~\bibnamefont {Wei}}, \bibinfo {author} {\bibfnamefont {S.}~\bibnamefont
  {Zhao}}, \bibinfo {author} {\bibfnamefont {W.}~\bibnamefont {Zhao}}, \bibinfo
  {author} {\bibfnamefont {Z.}~\bibnamefont {Zhang}}, \bibinfo {author}
  {\bibfnamefont {K.}~\bibnamefont {Yumigeta}}, \emph {et~al.},\ }\bibfield
  {title} {\bibinfo {title} {Mott and generalized wigner crystal states in
  wse2/ws2 moir{\'e} superlattices},\ }\href@noop {} {\bibfield  {journal}
  {\bibinfo  {journal} {Nature}\ }\textbf {\bibinfo {volume} {579}},\ \bibinfo
  {pages} {359} (\bibinfo {year} {2020})}\BibitemShut {NoStop}%
\bibitem [{\citenamefont {Xu}\ \emph {et~al.}(2020)\citenamefont {Xu},
  \citenamefont {Liu}, \citenamefont {Rhodes}, \citenamefont {Watanabe},
  \citenamefont {Taniguchi}, \citenamefont {Hone}, \citenamefont {Elser},
  \citenamefont {Mak},\ and\ \citenamefont {Shan}}]{xu2020correlated}%
  \BibitemOpen
  \bibfield  {author} {\bibinfo {author} {\bibfnamefont {Y.}~\bibnamefont
  {Xu}}, \bibinfo {author} {\bibfnamefont {S.}~\bibnamefont {Liu}}, \bibinfo
  {author} {\bibfnamefont {D.~A.}\ \bibnamefont {Rhodes}}, \bibinfo {author}
  {\bibfnamefont {K.}~\bibnamefont {Watanabe}}, \bibinfo {author}
  {\bibfnamefont {T.}~\bibnamefont {Taniguchi}}, \bibinfo {author}
  {\bibfnamefont {J.}~\bibnamefont {Hone}}, \bibinfo {author} {\bibfnamefont
  {V.}~\bibnamefont {Elser}}, \bibinfo {author} {\bibfnamefont {K.~F.}\
  \bibnamefont {Mak}},\ and\ \bibinfo {author} {\bibfnamefont {J.}~\bibnamefont
  {Shan}},\ }\bibfield  {title} {\bibinfo {title} {Correlated insulating states
  at fractional fillings of moir{\'e} superlattices},\ }\href@noop {}
  {\bibfield  {journal} {\bibinfo  {journal} {Nature}\ }\textbf {\bibinfo
  {volume} {587}},\ \bibinfo {pages} {214} (\bibinfo {year}
  {2020})}\BibitemShut {NoStop}%
\bibitem [{\citenamefont {Ghiotto}\ \emph {et~al.}(2021)\citenamefont
  {Ghiotto}, \citenamefont {Shih}, \citenamefont {Pereira}, \citenamefont
  {Rhodes}, \citenamefont {Kim}, \citenamefont {Zang}, \citenamefont {Millis},
  \citenamefont {Watanabe}, \citenamefont {Taniguchi}, \citenamefont {Hone}
  \emph {et~al.}}]{ghiotto2021quantum}%
  \BibitemOpen
  \bibfield  {author} {\bibinfo {author} {\bibfnamefont {A.}~\bibnamefont
  {Ghiotto}}, \bibinfo {author} {\bibfnamefont {E.-M.}\ \bibnamefont {Shih}},
  \bibinfo {author} {\bibfnamefont {G.~S.}\ \bibnamefont {Pereira}}, \bibinfo
  {author} {\bibfnamefont {D.~A.}\ \bibnamefont {Rhodes}}, \bibinfo {author}
  {\bibfnamefont {B.}~\bibnamefont {Kim}}, \bibinfo {author} {\bibfnamefont
  {J.}~\bibnamefont {Zang}}, \bibinfo {author} {\bibfnamefont {A.~J.}\
  \bibnamefont {Millis}}, \bibinfo {author} {\bibfnamefont {K.}~\bibnamefont
  {Watanabe}}, \bibinfo {author} {\bibfnamefont {T.}~\bibnamefont {Taniguchi}},
  \bibinfo {author} {\bibfnamefont {J.~C.}\ \bibnamefont {Hone}}, \emph
  {et~al.},\ }\bibfield  {title} {\bibinfo {title} {Quantum criticality in
  twisted transition metal dichalcogenides},\ }\href@noop {} {\bibfield
  {journal} {\bibinfo  {journal} {Nature}\ }\textbf {\bibinfo {volume} {597}},\
  \bibinfo {pages} {345} (\bibinfo {year} {2021})}\BibitemShut {NoStop}%
\bibitem [{\citenamefont {Li}\ \emph {et~al.}(2021{\natexlab{a}})\citenamefont
  {Li}, \citenamefont {Jiang}, \citenamefont {Shen}, \citenamefont {Zhang},
  \citenamefont {Li}, \citenamefont {Tao}, \citenamefont {Devakul},
  \citenamefont {Watanabe}, \citenamefont {Taniguchi}, \citenamefont {Fu} \emph
  {et~al.}}]{li2021quantum}%
  \BibitemOpen
  \bibfield  {author} {\bibinfo {author} {\bibfnamefont {T.}~\bibnamefont
  {Li}}, \bibinfo {author} {\bibfnamefont {S.}~\bibnamefont {Jiang}}, \bibinfo
  {author} {\bibfnamefont {B.}~\bibnamefont {Shen}}, \bibinfo {author}
  {\bibfnamefont {Y.}~\bibnamefont {Zhang}}, \bibinfo {author} {\bibfnamefont
  {L.}~\bibnamefont {Li}}, \bibinfo {author} {\bibfnamefont {Z.}~\bibnamefont
  {Tao}}, \bibinfo {author} {\bibfnamefont {T.}~\bibnamefont {Devakul}},
  \bibinfo {author} {\bibfnamefont {K.}~\bibnamefont {Watanabe}}, \bibinfo
  {author} {\bibfnamefont {T.}~\bibnamefont {Taniguchi}}, \bibinfo {author}
  {\bibfnamefont {L.}~\bibnamefont {Fu}}, \emph {et~al.},\ }\bibfield  {title}
  {\bibinfo {title} {{Quantum anomalous Hall effect from intertwined moir{\'e}
  bands}},\ }\href@noop {} {\bibfield  {journal} {\bibinfo  {journal} {Nature}\
  }\textbf {\bibinfo {volume} {600}},\ \bibinfo {pages} {641} (\bibinfo {year}
  {2021}{\natexlab{a}})}\BibitemShut {NoStop}%
\bibitem [{\citenamefont {Tao}\ \emph {et~al.}(2022)\citenamefont {Tao},
  \citenamefont {Shen}, \citenamefont {Jiang}, \citenamefont {Li},
  \citenamefont {Li}, \citenamefont {Ma}, \citenamefont {Zhao}, \citenamefont
  {Hu}, \citenamefont {Pistunova}, \citenamefont {Watanabe} \emph
  {et~al.}}]{tao2022valley}%
  \BibitemOpen
  \bibfield  {author} {\bibinfo {author} {\bibfnamefont {Z.}~\bibnamefont
  {Tao}}, \bibinfo {author} {\bibfnamefont {B.}~\bibnamefont {Shen}}, \bibinfo
  {author} {\bibfnamefont {S.}~\bibnamefont {Jiang}}, \bibinfo {author}
  {\bibfnamefont {T.}~\bibnamefont {Li}}, \bibinfo {author} {\bibfnamefont
  {L.}~\bibnamefont {Li}}, \bibinfo {author} {\bibfnamefont {L.}~\bibnamefont
  {Ma}}, \bibinfo {author} {\bibfnamefont {W.}~\bibnamefont {Zhao}}, \bibinfo
  {author} {\bibfnamefont {J.}~\bibnamefont {Hu}}, \bibinfo {author}
  {\bibfnamefont {K.}~\bibnamefont {Pistunova}}, \bibinfo {author}
  {\bibfnamefont {K.}~\bibnamefont {Watanabe}}, \emph {et~al.},\ }\bibfield
  {title} {\bibinfo {title} {{Valley-coherent quantum anomalous Hall state in
  AB-stacked MoTe$_2$/WSe$_2$ bilayers}},\ }\href@noop {} {\bibfield  {journal}
  {\bibinfo  {journal} {arXiv preprint arXiv:2208.07452}\ } (\bibinfo {year}
  {2022})}\BibitemShut {NoStop}%
\bibitem [{\citenamefont {Mak}\ and\ \citenamefont
  {Shan}(2022)}]{mak2022semiconductor}%
  \BibitemOpen
  \bibfield  {author} {\bibinfo {author} {\bibfnamefont {K.~F.}\ \bibnamefont
  {Mak}}\ and\ \bibinfo {author} {\bibfnamefont {J.}~\bibnamefont {Shan}},\
  }\bibfield  {title} {\bibinfo {title} {Semiconductor moir{\'e} materials},\
  }\href@noop {} {\bibfield  {journal} {\bibinfo  {journal} {Nature
  Nanotechnology}\ }\textbf {\bibinfo {volume} {17}},\ \bibinfo {pages} {686}
  (\bibinfo {year} {2022})}\BibitemShut {NoStop}%
\bibitem [{\citenamefont {Zeng}\ \emph {et~al.}(2023)\citenamefont {Zeng},
  \citenamefont {Xia}, \citenamefont {Kang}, \citenamefont {Zhu}, \citenamefont
  {Kn{\"u}ppel}, \citenamefont {Vaswani}, \citenamefont {Watanabe},
  \citenamefont {Taniguchi}, \citenamefont {Mak},\ and\ \citenamefont
  {Shan}}]{zeng2023thermodynamic}%
  \BibitemOpen
  \bibfield  {author} {\bibinfo {author} {\bibfnamefont {Y.}~\bibnamefont
  {Zeng}}, \bibinfo {author} {\bibfnamefont {Z.}~\bibnamefont {Xia}}, \bibinfo
  {author} {\bibfnamefont {K.}~\bibnamefont {Kang}}, \bibinfo {author}
  {\bibfnamefont {J.}~\bibnamefont {Zhu}}, \bibinfo {author} {\bibfnamefont
  {P.}~\bibnamefont {Kn{\"u}ppel}}, \bibinfo {author} {\bibfnamefont
  {C.}~\bibnamefont {Vaswani}}, \bibinfo {author} {\bibfnamefont
  {K.}~\bibnamefont {Watanabe}}, \bibinfo {author} {\bibfnamefont
  {T.}~\bibnamefont {Taniguchi}}, \bibinfo {author} {\bibfnamefont {K.~F.}\
  \bibnamefont {Mak}},\ and\ \bibinfo {author} {\bibfnamefont {J.}~\bibnamefont
  {Shan}},\ }\bibfield  {title} {\bibinfo {title} {{Thermodynamic evidence of
  fractional Chern insulator in moir{\'e} MoTe$_2$}},\ }\href@noop {}
  {\bibfield  {journal} {\bibinfo  {journal} {Nature}\ }\textbf {\bibinfo
  {volume} {622}},\ \bibinfo {pages} {69} (\bibinfo {year} {2023})}\BibitemShut
  {NoStop}%
\bibitem [{\citenamefont {Cai}\ \emph {et~al.}(2023)\citenamefont {Cai},
  \citenamefont {Anderson}, \citenamefont {Wang}, \citenamefont {Zhang},
  \citenamefont {Liu}, \citenamefont {Holtzmann}, \citenamefont {Zhang},
  \citenamefont {Fan}, \citenamefont {Taniguchi}, \citenamefont {Watanabe}
  \emph {et~al.}}]{cai2023signatures}%
  \BibitemOpen
  \bibfield  {author} {\bibinfo {author} {\bibfnamefont {J.}~\bibnamefont
  {Cai}}, \bibinfo {author} {\bibfnamefont {E.}~\bibnamefont {Anderson}},
  \bibinfo {author} {\bibfnamefont {C.}~\bibnamefont {Wang}}, \bibinfo {author}
  {\bibfnamefont {X.}~\bibnamefont {Zhang}}, \bibinfo {author} {\bibfnamefont
  {X.}~\bibnamefont {Liu}}, \bibinfo {author} {\bibfnamefont {W.}~\bibnamefont
  {Holtzmann}}, \bibinfo {author} {\bibfnamefont {Y.}~\bibnamefont {Zhang}},
  \bibinfo {author} {\bibfnamefont {F.}~\bibnamefont {Fan}}, \bibinfo {author}
  {\bibfnamefont {T.}~\bibnamefont {Taniguchi}}, \bibinfo {author}
  {\bibfnamefont {K.}~\bibnamefont {Watanabe}}, \emph {et~al.},\ }\bibfield
  {title} {\bibinfo {title} {{Signatures of Fractional Quantum Anomalous Hall
  States in Twisted MoTe$_2$}},\ }\href@noop {} {\bibfield  {journal} {\bibinfo
   {journal} {Nature}\ }\textbf {\bibinfo {volume} {622}},\ \bibinfo {pages}
  {63} (\bibinfo {year} {2023})}\BibitemShut {NoStop}%
\bibitem [{\citenamefont {Park}\ \emph {et~al.}(2023)\citenamefont {Park},
  \citenamefont {Cai}, \citenamefont {Anderson}, \citenamefont {Zhang},
  \citenamefont {Zhu}, \citenamefont {Liu}, \citenamefont {Wang}, \citenamefont
  {Holtzmann}, \citenamefont {Hu}, \citenamefont {Liu} \emph
  {et~al.}}]{park2023observation}%
  \BibitemOpen
  \bibfield  {author} {\bibinfo {author} {\bibfnamefont {H.}~\bibnamefont
  {Park}}, \bibinfo {author} {\bibfnamefont {J.}~\bibnamefont {Cai}}, \bibinfo
  {author} {\bibfnamefont {E.}~\bibnamefont {Anderson}}, \bibinfo {author}
  {\bibfnamefont {Y.}~\bibnamefont {Zhang}}, \bibinfo {author} {\bibfnamefont
  {J.}~\bibnamefont {Zhu}}, \bibinfo {author} {\bibfnamefont {X.}~\bibnamefont
  {Liu}}, \bibinfo {author} {\bibfnamefont {C.}~\bibnamefont {Wang}}, \bibinfo
  {author} {\bibfnamefont {W.}~\bibnamefont {Holtzmann}}, \bibinfo {author}
  {\bibfnamefont {C.}~\bibnamefont {Hu}}, \bibinfo {author} {\bibfnamefont
  {Z.}~\bibnamefont {Liu}}, \emph {et~al.},\ }\bibfield  {title} {\bibinfo
  {title} {{Observation of fractionally quantized anomalous Hall effect}},\
  }\href@noop {} {\bibfield  {journal} {\bibinfo  {journal} {Nature}\ }\textbf
  {\bibinfo {volume} {622}},\ \bibinfo {pages} {74} (\bibinfo {year}
  {2023})}\BibitemShut {NoStop}%
\bibitem [{\citenamefont {Xu}\ \emph {et~al.}(2023)\citenamefont {Xu},
  \citenamefont {Sun}, \citenamefont {Jia}, \citenamefont {Liu}, \citenamefont
  {Xu}, \citenamefont {Li}, \citenamefont {Gu}, \citenamefont {Watanabe},
  \citenamefont {Taniguchi}, \citenamefont {Tong}, \citenamefont {Jia},
  \citenamefont {Shi}, \citenamefont {Jiang}, \citenamefont {Zhang},
  \citenamefont {Liu},\ and\ \citenamefont {Li}}]{xu2023observation}%
  \BibitemOpen
  \bibfield  {author} {\bibinfo {author} {\bibfnamefont {F.}~\bibnamefont
  {Xu}}, \bibinfo {author} {\bibfnamefont {Z.}~\bibnamefont {Sun}}, \bibinfo
  {author} {\bibfnamefont {T.}~\bibnamefont {Jia}}, \bibinfo {author}
  {\bibfnamefont {C.}~\bibnamefont {Liu}}, \bibinfo {author} {\bibfnamefont
  {C.}~\bibnamefont {Xu}}, \bibinfo {author} {\bibfnamefont {C.}~\bibnamefont
  {Li}}, \bibinfo {author} {\bibfnamefont {Y.}~\bibnamefont {Gu}}, \bibinfo
  {author} {\bibfnamefont {K.}~\bibnamefont {Watanabe}}, \bibinfo {author}
  {\bibfnamefont {T.}~\bibnamefont {Taniguchi}}, \bibinfo {author}
  {\bibfnamefont {B.}~\bibnamefont {Tong}}, \bibinfo {author} {\bibfnamefont
  {J.}~\bibnamefont {Jia}}, \bibinfo {author} {\bibfnamefont {Z.}~\bibnamefont
  {Shi}}, \bibinfo {author} {\bibfnamefont {S.}~\bibnamefont {Jiang}}, \bibinfo
  {author} {\bibfnamefont {Y.}~\bibnamefont {Zhang}}, \bibinfo {author}
  {\bibfnamefont {X.}~\bibnamefont {Liu}},\ and\ \bibinfo {author}
  {\bibfnamefont {T.}~\bibnamefont {Li}},\ }\bibfield  {title} {\bibinfo
  {title} {Observation of integer and fractional quantum anomalous hall effects
  in twisted bilayer ${\mathrm{mote}}_{2}$},\ }\href
  {https://doi.org/10.1103/PhysRevX.13.031037} {\bibfield  {journal} {\bibinfo
  {journal} {Phys. Rev. X}\ }\textbf {\bibinfo {volume} {13}},\ \bibinfo
  {pages} {031037} (\bibinfo {year} {2023})}\BibitemShut {NoStop}%
\bibitem [{\citenamefont {Kennes}\ \emph {et~al.}(2021)\citenamefont {Kennes},
  \citenamefont {Claassen}, \citenamefont {Xian}, \citenamefont {Georges},
  \citenamefont {Millis}, \citenamefont {Hone}, \citenamefont {Dean},
  \citenamefont {Basov}, \citenamefont {Pasupathy},\ and\ \citenamefont
  {Rubio}}]{kennes2021moire}%
  \BibitemOpen
  \bibfield  {author} {\bibinfo {author} {\bibfnamefont {D.~M.}\ \bibnamefont
  {Kennes}}, \bibinfo {author} {\bibfnamefont {M.}~\bibnamefont {Claassen}},
  \bibinfo {author} {\bibfnamefont {L.}~\bibnamefont {Xian}}, \bibinfo {author}
  {\bibfnamefont {A.}~\bibnamefont {Georges}}, \bibinfo {author} {\bibfnamefont
  {A.~J.}\ \bibnamefont {Millis}}, \bibinfo {author} {\bibfnamefont
  {J.}~\bibnamefont {Hone}}, \bibinfo {author} {\bibfnamefont {C.~R.}\
  \bibnamefont {Dean}}, \bibinfo {author} {\bibfnamefont {D.}~\bibnamefont
  {Basov}}, \bibinfo {author} {\bibfnamefont {A.~N.}\ \bibnamefont
  {Pasupathy}},\ and\ \bibinfo {author} {\bibfnamefont {A.}~\bibnamefont
  {Rubio}},\ }\bibfield  {title} {\bibinfo {title} {Moir{\'e} heterostructures
  as a condensed-matter quantum simulator},\ }\href@noop {} {\bibfield
  {journal} {\bibinfo  {journal} {Nature Physics}\ }\textbf {\bibinfo {volume}
  {17}},\ \bibinfo {pages} {155} (\bibinfo {year} {2021})}\BibitemShut
  {NoStop}%
\bibitem [{\citenamefont {Ghorashi}\ \emph {et~al.}(2023)\citenamefont
  {Ghorashi}, \citenamefont {Dunbrack}, \citenamefont {Abouelkomsan},
  \citenamefont {Sun}, \citenamefont {Du},\ and\ \citenamefont
  {Cano}}]{ghorashi2023topological}%
  \BibitemOpen
  \bibfield  {author} {\bibinfo {author} {\bibfnamefont {S.~A.~A.}\
  \bibnamefont {Ghorashi}}, \bibinfo {author} {\bibfnamefont {A.}~\bibnamefont
  {Dunbrack}}, \bibinfo {author} {\bibfnamefont {A.}~\bibnamefont
  {Abouelkomsan}}, \bibinfo {author} {\bibfnamefont {J.}~\bibnamefont {Sun}},
  \bibinfo {author} {\bibfnamefont {X.}~\bibnamefont {Du}},\ and\ \bibinfo
  {author} {\bibfnamefont {J.}~\bibnamefont {Cano}},\ }\bibfield  {title}
  {\bibinfo {title} {Topological and stacked flat bands in bilayer graphene
  with a superlattice potential},\ }\href
  {https://doi.org/10.1103/PhysRevLett.130.196201} {\bibfield  {journal}
  {\bibinfo  {journal} {Phys. Rev. Lett.}\ }\textbf {\bibinfo {volume} {130}},\
  \bibinfo {pages} {196201} (\bibinfo {year} {2023})}\BibitemShut {NoStop}%
\bibitem [{\citenamefont {Ghorashi}\ and\ \citenamefont
  {Cano}(2023)}]{ghorashi2023multilayer}%
  \BibitemOpen
  \bibfield  {author} {\bibinfo {author} {\bibfnamefont {S.~A.~A.}\
  \bibnamefont {Ghorashi}}\ and\ \bibinfo {author} {\bibfnamefont
  {J.}~\bibnamefont {Cano}},\ }\bibfield  {title} {\bibinfo {title} {Multilayer
  graphene with a superlattice potential},\ }\href
  {https://doi.org/10.1103/PhysRevB.107.195423} {\bibfield  {journal} {\bibinfo
   {journal} {Phys. Rev. B}\ }\textbf {\bibinfo {volume} {107}},\ \bibinfo
  {pages} {195423} (\bibinfo {year} {2023})}\BibitemShut {NoStop}%
\bibitem [{\citenamefont {Cr{\'e}pel}\ \emph {et~al.}(2023)\citenamefont
  {Cr{\'e}pel}, \citenamefont {Dunbrack}, \citenamefont {Guerci}, \citenamefont
  {Bonini},\ and\ \citenamefont {Cano}}]{crepel2023chiral}%
  \BibitemOpen
  \bibfield  {author} {\bibinfo {author} {\bibfnamefont {V.}~\bibnamefont
  {Cr{\'e}pel}}, \bibinfo {author} {\bibfnamefont {A.}~\bibnamefont
  {Dunbrack}}, \bibinfo {author} {\bibfnamefont {D.}~\bibnamefont {Guerci}},
  \bibinfo {author} {\bibfnamefont {J.}~\bibnamefont {Bonini}},\ and\ \bibinfo
  {author} {\bibfnamefont {J.}~\bibnamefont {Cano}},\ }\bibfield  {title}
  {\bibinfo {title} {Chiral model of twisted bilayer graphene realized in a
  monolayer},\ }\href@noop {} {\bibfield  {journal} {\bibinfo  {journal} {arXiv
  preprint arXiv:2305.14423}\ } (\bibinfo {year} {2023})}\BibitemShut {NoStop}%
\bibitem [{\citenamefont {Gao}\ \emph {et~al.}(2023)\citenamefont {Gao},
  \citenamefont {Dong}, \citenamefont {Ledwith}, \citenamefont {Parker},\ and\
  \citenamefont {Khalaf}}]{gao2023untwisting}%
  \BibitemOpen
  \bibfield  {author} {\bibinfo {author} {\bibfnamefont {Q.}~\bibnamefont
  {Gao}}, \bibinfo {author} {\bibfnamefont {J.}~\bibnamefont {Dong}}, \bibinfo
  {author} {\bibfnamefont {P.}~\bibnamefont {Ledwith}}, \bibinfo {author}
  {\bibfnamefont {D.}~\bibnamefont {Parker}},\ and\ \bibinfo {author}
  {\bibfnamefont {E.}~\bibnamefont {Khalaf}},\ }\bibfield  {title} {\bibinfo
  {title} {Untwisting moir\'e physics: Almost ideal bands and fractional chern
  insulators in periodically strained monolayer graphene},\ }\href
  {https://doi.org/10.1103/PhysRevLett.131.096401} {\bibfield  {journal}
  {\bibinfo  {journal} {Phys. Rev. Lett.}\ }\textbf {\bibinfo {volume} {131}},\
  \bibinfo {pages} {096401} (\bibinfo {year} {2023})}\BibitemShut {NoStop}%
\bibitem [{\citenamefont {Wan}\ \emph {et~al.}(2023)\citenamefont {Wan},
  \citenamefont {Sarkar}, \citenamefont {Lin},\ and\ \citenamefont
  {Sun}}]{wan2023topological}%
  \BibitemOpen
  \bibfield  {author} {\bibinfo {author} {\bibfnamefont {X.}~\bibnamefont
  {Wan}}, \bibinfo {author} {\bibfnamefont {S.}~\bibnamefont {Sarkar}},
  \bibinfo {author} {\bibfnamefont {S.-Z.}\ \bibnamefont {Lin}},\ and\ \bibinfo
  {author} {\bibfnamefont {K.}~\bibnamefont {Sun}},\ }\bibfield  {title}
  {\bibinfo {title} {Topological exact flat bands in two-dimensional materials
  under periodic strain},\ }\href@noop {} {\bibfield  {journal} {\bibinfo
  {journal} {Physical Review Letters}\ }\textbf {\bibinfo {volume} {130}},\
  \bibinfo {pages} {216401} (\bibinfo {year} {2023})}\BibitemShut {NoStop}%
\bibitem [{\citenamefont {Tarnopolsky}\ \emph {et~al.}(2019)\citenamefont
  {Tarnopolsky}, \citenamefont {Kruchkov},\ and\ \citenamefont
  {Vishwanath}}]{tarnopolsky2019origin}%
  \BibitemOpen
  \bibfield  {author} {\bibinfo {author} {\bibfnamefont {G.}~\bibnamefont
  {Tarnopolsky}}, \bibinfo {author} {\bibfnamefont {A.~J.}\ \bibnamefont
  {Kruchkov}},\ and\ \bibinfo {author} {\bibfnamefont {A.}~\bibnamefont
  {Vishwanath}},\ }\bibfield  {title} {\bibinfo {title} {Origin of magic angles
  in twisted bilayer graphene},\ }\href
  {https://doi.org/10.1103/PhysRevLett.122.106405} {\bibfield  {journal}
  {\bibinfo  {journal} {Phys. Rev. Lett.}\ }\textbf {\bibinfo {volume} {122}},\
  \bibinfo {pages} {106405} (\bibinfo {year} {2019})}\BibitemShut {NoStop}%
\bibitem [{\citenamefont {Liu}\ \emph {et~al.}(2019)\citenamefont {Liu},
  \citenamefont {Liu},\ and\ \citenamefont {Dai}}]{liu2019pseudo}%
  \BibitemOpen
  \bibfield  {author} {\bibinfo {author} {\bibfnamefont {J.}~\bibnamefont
  {Liu}}, \bibinfo {author} {\bibfnamefont {J.}~\bibnamefont {Liu}},\ and\
  \bibinfo {author} {\bibfnamefont {X.}~\bibnamefont {Dai}},\ }\bibfield
  {title} {\bibinfo {title} {Pseudo landau level representation of twisted
  bilayer graphene: Band topology and implications on the correlated insulating
  phase},\ }\href {https://doi.org/10.1103/PhysRevB.99.155415} {\bibfield
  {journal} {\bibinfo  {journal} {Phys. Rev. B}\ }\textbf {\bibinfo {volume}
  {99}},\ \bibinfo {pages} {155415} (\bibinfo {year} {2019})}\BibitemShut
  {NoStop}%
\bibitem [{\citenamefont {Bultinck}\ \emph {et~al.}(2020)\citenamefont
  {Bultinck}, \citenamefont {Chatterjee},\ and\ \citenamefont
  {Zaletel}}]{bultinck2020mechanism}%
  \BibitemOpen
  \bibfield  {author} {\bibinfo {author} {\bibfnamefont {N.}~\bibnamefont
  {Bultinck}}, \bibinfo {author} {\bibfnamefont {S.}~\bibnamefont
  {Chatterjee}},\ and\ \bibinfo {author} {\bibfnamefont {M.~P.}\ \bibnamefont
  {Zaletel}},\ }\bibfield  {title} {\bibinfo {title} {Mechanism for anomalous
  hall ferromagnetism in twisted bilayer graphene},\ }\href
  {https://doi.org/10.1103/PhysRevLett.124.166601} {\bibfield  {journal}
  {\bibinfo  {journal} {Phys. Rev. Lett.}\ }\textbf {\bibinfo {volume} {124}},\
  \bibinfo {pages} {166601} (\bibinfo {year} {2020})}\BibitemShut {NoStop}%
\bibitem [{\citenamefont {Ledwith}\ \emph {et~al.}(2021)\citenamefont
  {Ledwith}, \citenamefont {Khalaf},\ and\ \citenamefont
  {Vishwanath}}]{ledwith2021strong}%
  \BibitemOpen
  \bibfield  {author} {\bibinfo {author} {\bibfnamefont {P.~J.}\ \bibnamefont
  {Ledwith}}, \bibinfo {author} {\bibfnamefont {E.}~\bibnamefont {Khalaf}},\
  and\ \bibinfo {author} {\bibfnamefont {A.}~\bibnamefont {Vishwanath}},\
  }\bibfield  {title} {\bibinfo {title} {Strong coupling theory of magic-angle
  graphene: A pedagogical introduction},\ }\href@noop {} {\bibfield  {journal}
  {\bibinfo  {journal} {Annals of Physics}\ }\textbf {\bibinfo {volume}
  {435}},\ \bibinfo {pages} {168646} (\bibinfo {year} {2021})}\BibitemShut
  {NoStop}%
\bibitem [{\citenamefont {Wang}\ \emph {et~al.}(2021)\citenamefont {Wang},
  \citenamefont {Cano}, \citenamefont {Millis}, \citenamefont {Liu},\ and\
  \citenamefont {Yang}}]{wang2021exact}%
  \BibitemOpen
  \bibfield  {author} {\bibinfo {author} {\bibfnamefont {J.}~\bibnamefont
  {Wang}}, \bibinfo {author} {\bibfnamefont {J.}~\bibnamefont {Cano}}, \bibinfo
  {author} {\bibfnamefont {A.~J.}\ \bibnamefont {Millis}}, \bibinfo {author}
  {\bibfnamefont {Z.}~\bibnamefont {Liu}},\ and\ \bibinfo {author}
  {\bibfnamefont {B.}~\bibnamefont {Yang}},\ }\bibfield  {title} {\bibinfo
  {title} {Exact landau level description of geometry and interaction in a
  flatband},\ }\href {https://doi.org/10.1103/PhysRevLett.127.246403}
  {\bibfield  {journal} {\bibinfo  {journal} {Phys. Rev. Lett.}\ }\textbf
  {\bibinfo {volume} {127}},\ \bibinfo {pages} {246403} (\bibinfo {year}
  {2021})}\BibitemShut {NoStop}%
\bibitem [{\citenamefont {Khalaf}\ \emph {et~al.}(2021)\citenamefont {Khalaf},
  \citenamefont {Chatterjee}, \citenamefont {Bultinck}, \citenamefont
  {Zaletel},\ and\ \citenamefont {Vishwanath}}]{khalaf2021charged}%
  \BibitemOpen
  \bibfield  {author} {\bibinfo {author} {\bibfnamefont {E.}~\bibnamefont
  {Khalaf}}, \bibinfo {author} {\bibfnamefont {S.}~\bibnamefont {Chatterjee}},
  \bibinfo {author} {\bibfnamefont {N.}~\bibnamefont {Bultinck}}, \bibinfo
  {author} {\bibfnamefont {M.~P.}\ \bibnamefont {Zaletel}},\ and\ \bibinfo
  {author} {\bibfnamefont {A.}~\bibnamefont {Vishwanath}},\ }\bibfield  {title}
  {\bibinfo {title} {Charged skyrmions and topological origin of
  superconductivity in magic-angle graphene},\ }\href@noop {} {\bibfield
  {journal} {\bibinfo  {journal} {Science advances}\ }\textbf {\bibinfo
  {volume} {7}},\ \bibinfo {pages} {eabf5299} (\bibinfo {year}
  {2021})}\BibitemShut {NoStop}%
\bibitem [{\citenamefont {Chatterjee}\ \emph {et~al.}(2022)\citenamefont
  {Chatterjee}, \citenamefont {Ippoliti},\ and\ \citenamefont
  {Zaletel}}]{chatterjee2022skyrmion}%
  \BibitemOpen
  \bibfield  {author} {\bibinfo {author} {\bibfnamefont {S.}~\bibnamefont
  {Chatterjee}}, \bibinfo {author} {\bibfnamefont {M.}~\bibnamefont
  {Ippoliti}},\ and\ \bibinfo {author} {\bibfnamefont {M.~P.}\ \bibnamefont
  {Zaletel}},\ }\bibfield  {title} {\bibinfo {title} {{Skyrmion
  superconductivity: DMRG evidence for a topological route to
  superconductivity}},\ }\href {https://doi.org/10.1103/PhysRevB.106.035421}
  {\bibfield  {journal} {\bibinfo  {journal} {Phys. Rev. B}\ }\textbf {\bibinfo
  {volume} {106}},\ \bibinfo {pages} {035421} (\bibinfo {year}
  {2022})}\BibitemShut {NoStop}%
\bibitem [{\citenamefont {T{\"o}rm{\"a}}\ \emph {et~al.}(2022)\citenamefont
  {T{\"o}rm{\"a}}, \citenamefont {Peotta},\ and\ \citenamefont
  {Bernevig}}]{torma2022superconductivity}%
  \BibitemOpen
  \bibfield  {author} {\bibinfo {author} {\bibfnamefont {P.}~\bibnamefont
  {T{\"o}rm{\"a}}}, \bibinfo {author} {\bibfnamefont {S.}~\bibnamefont
  {Peotta}},\ and\ \bibinfo {author} {\bibfnamefont {B.~A.}\ \bibnamefont
  {Bernevig}},\ }\bibfield  {title} {\bibinfo {title} {Superconductivity,
  superfluidity and quantum geometry in twisted multilayer systems},\
  }\href@noop {} {\bibfield  {journal} {\bibinfo  {journal} {Nature Reviews
  Physics}\ }\textbf {\bibinfo {volume} {4}},\ \bibinfo {pages} {528} (\bibinfo
  {year} {2022})}\BibitemShut {NoStop}%
\bibitem [{\citenamefont {Forsythe}\ \emph {et~al.}(2018)\citenamefont
  {Forsythe}, \citenamefont {Zhou}, \citenamefont {Watanabe}, \citenamefont
  {Taniguchi}, \citenamefont {Pasupathy}, \citenamefont {Moon}, \citenamefont
  {Koshino}, \citenamefont {Kim},\ and\ \citenamefont
  {Dean}}]{forsythe2018band}%
  \BibitemOpen
  \bibfield  {author} {\bibinfo {author} {\bibfnamefont {C.}~\bibnamefont
  {Forsythe}}, \bibinfo {author} {\bibfnamefont {X.}~\bibnamefont {Zhou}},
  \bibinfo {author} {\bibfnamefont {K.}~\bibnamefont {Watanabe}}, \bibinfo
  {author} {\bibfnamefont {T.}~\bibnamefont {Taniguchi}}, \bibinfo {author}
  {\bibfnamefont {A.}~\bibnamefont {Pasupathy}}, \bibinfo {author}
  {\bibfnamefont {P.}~\bibnamefont {Moon}}, \bibinfo {author} {\bibfnamefont
  {M.}~\bibnamefont {Koshino}}, \bibinfo {author} {\bibfnamefont
  {P.}~\bibnamefont {Kim}},\ and\ \bibinfo {author} {\bibfnamefont {C.~R.}\
  \bibnamefont {Dean}},\ }\bibfield  {title} {\bibinfo {title} {Band structure
  engineering of 2d materials using patterned dielectric superlattices},\
  }\href@noop {} {\bibfield  {journal} {\bibinfo  {journal} {Nature
  nanotechnology}\ }\textbf {\bibinfo {volume} {13}},\ \bibinfo {pages} {566}
  (\bibinfo {year} {2018})}\BibitemShut {NoStop}%
\bibitem [{\citenamefont {Li}\ \emph {et~al.}(2021{\natexlab{b}})\citenamefont
  {Li}, \citenamefont {Dietrich}, \citenamefont {Forsythe}, \citenamefont
  {Taniguchi}, \citenamefont {Watanabe}, \citenamefont {Moon},\ and\
  \citenamefont {Dean}}]{li2021anisotropic}%
  \BibitemOpen
  \bibfield  {author} {\bibinfo {author} {\bibfnamefont {Y.}~\bibnamefont
  {Li}}, \bibinfo {author} {\bibfnamefont {S.}~\bibnamefont {Dietrich}},
  \bibinfo {author} {\bibfnamefont {C.}~\bibnamefont {Forsythe}}, \bibinfo
  {author} {\bibfnamefont {T.}~\bibnamefont {Taniguchi}}, \bibinfo {author}
  {\bibfnamefont {K.}~\bibnamefont {Watanabe}}, \bibinfo {author}
  {\bibfnamefont {P.}~\bibnamefont {Moon}},\ and\ \bibinfo {author}
  {\bibfnamefont {C.~R.}\ \bibnamefont {Dean}},\ }\bibfield  {title} {\bibinfo
  {title} {Anisotropic band flattening in graphene with one-dimensional
  superlattices},\ }\href@noop {} {\bibfield  {journal} {\bibinfo  {journal}
  {Nature Nanotechnology}\ }\textbf {\bibinfo {volume} {16}},\ \bibinfo {pages}
  {525} (\bibinfo {year} {2021}{\natexlab{b}})}\BibitemShut {NoStop}%
\bibitem [{\citenamefont {Barcons~Ruiz}\ \emph {et~al.}(2022)\citenamefont
  {Barcons~Ruiz}, \citenamefont {Herzig~Sheinfux}, \citenamefont {Hoffmann},
  \citenamefont {Torre}, \citenamefont {Agarwal}, \citenamefont {Kumar},
  \citenamefont {Vistoli}, \citenamefont {Taniguchi}, \citenamefont {Watanabe},
  \citenamefont {Bachtold},\ and\ \citenamefont
  {Koppens}}]{barcons2022engineering}%
  \BibitemOpen
  \bibfield  {author} {\bibinfo {author} {\bibfnamefont {D.}~\bibnamefont
  {Barcons~Ruiz}}, \bibinfo {author} {\bibfnamefont {H.}~\bibnamefont
  {Herzig~Sheinfux}}, \bibinfo {author} {\bibfnamefont {R.}~\bibnamefont
  {Hoffmann}}, \bibinfo {author} {\bibfnamefont {I.}~\bibnamefont {Torre}},
  \bibinfo {author} {\bibfnamefont {H.}~\bibnamefont {Agarwal}}, \bibinfo
  {author} {\bibfnamefont {R.~K.}\ \bibnamefont {Kumar}}, \bibinfo {author}
  {\bibfnamefont {L.}~\bibnamefont {Vistoli}}, \bibinfo {author} {\bibfnamefont
  {T.}~\bibnamefont {Taniguchi}}, \bibinfo {author} {\bibfnamefont
  {K.}~\bibnamefont {Watanabe}}, \bibinfo {author} {\bibfnamefont
  {A.}~\bibnamefont {Bachtold}},\ and\ \bibinfo {author} {\bibfnamefont
  {F.~H.~L.}\ \bibnamefont {Koppens}},\ }\bibfield  {title} {\bibinfo {title}
  {Engineering high quality graphene superlattices via ion milled ultra-thin
  etching masks},\ }\href@noop {} {\bibfield  {journal} {\bibinfo  {journal}
  {Nature Communications}\ }\textbf {\bibinfo {volume} {13}},\ \bibinfo {pages}
  {6926} (\bibinfo {year} {2022})}\BibitemShut {NoStop}%
\bibitem [{\citenamefont {Sun}\ \emph {et~al.}(2023)\citenamefont {Sun},
  \citenamefont {Ghorashi}, \citenamefont {Watanabe}, \citenamefont
  {Taniguchi}, \citenamefont {Camino}, \citenamefont {Cano},\ and\
  \citenamefont {Du}}]{sun2023signature}%
  \BibitemOpen
  \bibfield  {author} {\bibinfo {author} {\bibfnamefont {J.}~\bibnamefont
  {Sun}}, \bibinfo {author} {\bibfnamefont {S.~A.~A.}\ \bibnamefont
  {Ghorashi}}, \bibinfo {author} {\bibfnamefont {K.}~\bibnamefont {Watanabe}},
  \bibinfo {author} {\bibfnamefont {T.}~\bibnamefont {Taniguchi}}, \bibinfo
  {author} {\bibfnamefont {F.}~\bibnamefont {Camino}}, \bibinfo {author}
  {\bibfnamefont {J.}~\bibnamefont {Cano}},\ and\ \bibinfo {author}
  {\bibfnamefont {X.}~\bibnamefont {Du}},\ }\bibfield  {title} {\bibinfo
  {title} {Signature of correlated insulator in electric field controlled
  superlattice},\ }\href@noop {} {\bibfield  {journal} {\bibinfo  {journal}
  {arXiv preprint arXiv:2306.06848}\ } (\bibinfo {year} {2023})}\BibitemShut
  {NoStop}%
\bibitem [{\citenamefont {Yasuda}\ \emph {et~al.}(2021)\citenamefont {Yasuda},
  \citenamefont {Wang}, \citenamefont {Watanabe}, \citenamefont {Taniguchi},\
  and\ \citenamefont {Jarillo-Herrero}}]{yasuda2021stacking}%
  \BibitemOpen
  \bibfield  {author} {\bibinfo {author} {\bibfnamefont {K.}~\bibnamefont
  {Yasuda}}, \bibinfo {author} {\bibfnamefont {X.}~\bibnamefont {Wang}},
  \bibinfo {author} {\bibfnamefont {K.}~\bibnamefont {Watanabe}}, \bibinfo
  {author} {\bibfnamefont {T.}~\bibnamefont {Taniguchi}},\ and\ \bibinfo
  {author} {\bibfnamefont {P.}~\bibnamefont {Jarillo-Herrero}},\ }\bibfield
  {title} {\bibinfo {title} {Stacking-engineered ferroelectricity in bilayer
  boron nitride},\ }\href@noop {} {\bibfield  {journal} {\bibinfo  {journal}
  {Science}\ }\textbf {\bibinfo {volume} {372}},\ \bibinfo {pages} {1458}
  (\bibinfo {year} {2021})}\BibitemShut {NoStop}%
\bibitem [{\citenamefont {Vizner~Stern}\ \emph {et~al.}(2021)\citenamefont
  {Vizner~Stern}, \citenamefont {Waschitz}, \citenamefont {Cao}, \citenamefont
  {Nevo}, \citenamefont {Watanabe}, \citenamefont {Taniguchi}, \citenamefont
  {Sela}, \citenamefont {Urbakh}, \citenamefont {Hod},\ and\ \citenamefont
  {Ben~Shalom}}]{vizner2021interfacial}%
  \BibitemOpen
  \bibfield  {author} {\bibinfo {author} {\bibfnamefont {M.}~\bibnamefont
  {Vizner~Stern}}, \bibinfo {author} {\bibfnamefont {Y.}~\bibnamefont
  {Waschitz}}, \bibinfo {author} {\bibfnamefont {W.}~\bibnamefont {Cao}},
  \bibinfo {author} {\bibfnamefont {I.}~\bibnamefont {Nevo}}, \bibinfo {author}
  {\bibfnamefont {K.}~\bibnamefont {Watanabe}}, \bibinfo {author}
  {\bibfnamefont {T.}~\bibnamefont {Taniguchi}}, \bibinfo {author}
  {\bibfnamefont {E.}~\bibnamefont {Sela}}, \bibinfo {author} {\bibfnamefont
  {M.}~\bibnamefont {Urbakh}}, \bibinfo {author} {\bibfnamefont
  {O.}~\bibnamefont {Hod}},\ and\ \bibinfo {author} {\bibfnamefont
  {M.}~\bibnamefont {Ben~Shalom}},\ }\bibfield  {title} {\bibinfo {title}
  {Interfacial ferroelectricity by van der waals sliding},\ }\href@noop {}
  {\bibfield  {journal} {\bibinfo  {journal} {Science}\ }\textbf {\bibinfo
  {volume} {372}},\ \bibinfo {pages} {1462} (\bibinfo {year}
  {2021})}\BibitemShut {NoStop}%
\bibitem [{\citenamefont {Wang}\ \emph
  {et~al.}(2022{\natexlab{a}})\citenamefont {Wang}, \citenamefont {Yasuda},
  \citenamefont {Zhang}, \citenamefont {Liu}, \citenamefont {Watanabe},
  \citenamefont {Taniguchi}, \citenamefont {Hone}, \citenamefont {Fu},\ and\
  \citenamefont {Jarillo-Herrero}}]{wang2022interfacial}%
  \BibitemOpen
  \bibfield  {author} {\bibinfo {author} {\bibfnamefont {X.}~\bibnamefont
  {Wang}}, \bibinfo {author} {\bibfnamefont {K.}~\bibnamefont {Yasuda}},
  \bibinfo {author} {\bibfnamefont {Y.}~\bibnamefont {Zhang}}, \bibinfo
  {author} {\bibfnamefont {S.}~\bibnamefont {Liu}}, \bibinfo {author}
  {\bibfnamefont {K.}~\bibnamefont {Watanabe}}, \bibinfo {author}
  {\bibfnamefont {T.}~\bibnamefont {Taniguchi}}, \bibinfo {author}
  {\bibfnamefont {J.}~\bibnamefont {Hone}}, \bibinfo {author} {\bibfnamefont
  {L.}~\bibnamefont {Fu}},\ and\ \bibinfo {author} {\bibfnamefont
  {P.}~\bibnamefont {Jarillo-Herrero}},\ }\bibfield  {title} {\bibinfo {title}
  {Interfacial ferroelectricity in rhombohedral-stacked bilayer transition
  metal dichalcogenides},\ }\href@noop {} {\bibfield  {journal} {\bibinfo
  {journal} {Nature nanotechnology}\ }\textbf {\bibinfo {volume} {17}},\
  \bibinfo {pages} {367} (\bibinfo {year} {2022}{\natexlab{a}})}\BibitemShut
  {NoStop}%
\bibitem [{\citenamefont {Kim}\ \emph {et~al.}(2023)\citenamefont {Kim},
  \citenamefont {Dominguez}, \citenamefont {Mayorga-Luna}, \citenamefont {Ye},
  \citenamefont {Embley}, \citenamefont {Tan}, \citenamefont {Ni},
  \citenamefont {Liu}, \citenamefont {Ford}, \citenamefont {Gao} \emph
  {et~al.}}]{kim2023electrostatic}%
  \BibitemOpen
  \bibfield  {author} {\bibinfo {author} {\bibfnamefont {D.~S.}\ \bibnamefont
  {Kim}}, \bibinfo {author} {\bibfnamefont {R.~C.}\ \bibnamefont {Dominguez}},
  \bibinfo {author} {\bibfnamefont {R.}~\bibnamefont {Mayorga-Luna}}, \bibinfo
  {author} {\bibfnamefont {D.}~\bibnamefont {Ye}}, \bibinfo {author}
  {\bibfnamefont {J.}~\bibnamefont {Embley}}, \bibinfo {author} {\bibfnamefont
  {T.}~\bibnamefont {Tan}}, \bibinfo {author} {\bibfnamefont {Y.}~\bibnamefont
  {Ni}}, \bibinfo {author} {\bibfnamefont {Z.}~\bibnamefont {Liu}}, \bibinfo
  {author} {\bibfnamefont {M.}~\bibnamefont {Ford}}, \bibinfo {author}
  {\bibfnamefont {F.~Y.}\ \bibnamefont {Gao}}, \emph {et~al.},\ }\bibfield
  {title} {\bibinfo {title} {Electrostatic moir\'e potential from twisted-{hBN}
  layers},\ }\href@noop {} {\bibfield  {journal} {\bibinfo  {journal} {arXiv
  preprint arXiv:2306.07841}\ } (\bibinfo {year} {2023})}\BibitemShut {NoStop}%
\bibitem [{\citenamefont {Wang}\ \emph
  {et~al.}(2022{\natexlab{b}})\citenamefont {Wang}, \citenamefont {Zhang},
  \citenamefont {Xie}, \citenamefont {Zhao}, \citenamefont {Sanborn},
  \citenamefont {Wang}, \citenamefont {Kahn}, \citenamefont {Watanabe},
  \citenamefont {Taniguchi}, \citenamefont {Zettl},\ and\ \citenamefont
  {Crommie}}]{wang2022engineering}%
  \BibitemOpen
  \bibfield  {author} {\bibinfo {author} {\bibfnamefont {F.}~\bibnamefont
  {Wang}}, \bibinfo {author} {\bibfnamefont {Z.}~\bibnamefont {Zhang}},
  \bibinfo {author} {\bibfnamefont {J.}~\bibnamefont {Xie}}, \bibinfo {author}
  {\bibfnamefont {W.}~\bibnamefont {Zhao}}, \bibinfo {author} {\bibfnamefont
  {C.}~\bibnamefont {Sanborn}}, \bibinfo {author} {\bibfnamefont
  {S.}~\bibnamefont {Wang}}, \bibinfo {author} {\bibfnamefont {S.}~\bibnamefont
  {Kahn}}, \bibinfo {author} {\bibfnamefont {K.}~\bibnamefont {Watanabe}},
  \bibinfo {author} {\bibfnamefont {T.}~\bibnamefont {Taniguchi}}, \bibinfo
  {author} {\bibfnamefont {A.}~\bibnamefont {Zettl}},\ and\ \bibinfo {author}
  {\bibfnamefont {M.}~\bibnamefont {Crommie}},\ }\bibfield  {title} {\bibinfo
  {title} {{Engineering correlated insulators in bilayer graphene with a remote
  Coulomb superlattice}}\ }\href {https://doi.org/10.21203/rs.3.rs-2286082/v1}
  {10.21203/rs.3.rs-2286082/v1} (\bibinfo {year}
  {2022}{\natexlab{b}})\BibitemShut {NoStop}%
\bibitem [{\citenamefont {Rokni}\ and\ \citenamefont
  {Lu}(2017)}]{rokni2017layer}%
  \BibitemOpen
  \bibfield  {author} {\bibinfo {author} {\bibfnamefont {H.}~\bibnamefont
  {Rokni}}\ and\ \bibinfo {author} {\bibfnamefont {W.}~\bibnamefont {Lu}},\
  }\bibfield  {title} {\bibinfo {title} {Layer-by-layer insight into
  electrostatic charge distribution of few-layer graphene},\ }\href@noop {}
  {\bibfield  {journal} {\bibinfo  {journal} {Scientific reports}\ }\textbf
  {\bibinfo {volume} {7}},\ \bibinfo {pages} {42821} (\bibinfo {year}
  {2017})}\BibitemShut {NoStop}%
\bibitem [{\citenamefont {Silvestrov}\ and\ \citenamefont
  {Recher}(2017)}]{silvestrov2017wigner}%
  \BibitemOpen
  \bibfield  {author} {\bibinfo {author} {\bibfnamefont {P.~G.}\ \bibnamefont
  {Silvestrov}}\ and\ \bibinfo {author} {\bibfnamefont {P.}~\bibnamefont
  {Recher}},\ }\bibfield  {title} {\bibinfo {title} {Wigner crystal phases in
  bilayer graphene},\ }\href {https://doi.org/10.1103/PhysRevB.95.075438}
  {\bibfield  {journal} {\bibinfo  {journal} {Phys. Rev. B}\ }\textbf {\bibinfo
  {volume} {95}},\ \bibinfo {pages} {075438} (\bibinfo {year}
  {2017})}\BibitemShut {NoStop}%
\bibitem [{\citenamefont {Joy}\ and\ \citenamefont
  {Skinner}(2023)}]{joy2023wigner}%
  \BibitemOpen
  \bibfield  {author} {\bibinfo {author} {\bibfnamefont {S.}~\bibnamefont
  {Joy}}\ and\ \bibinfo {author} {\bibfnamefont {B.}~\bibnamefont {Skinner}},\
  }\bibfield  {title} {\bibinfo {title} {Wigner crystallization in bernal
  bilayer graphene},\ }\href@noop {} {\bibfield  {journal} {\bibinfo  {journal}
  {arXiv preprint arXiv:2310.07751}\ } (\bibinfo {year} {2023})}\BibitemShut
  {NoStop}%
\bibitem [{\citenamefont {Su}\ \emph {et~al.}(2022)\citenamefont {Su},
  \citenamefont {Li}, \citenamefont {Zhang}, \citenamefont {Sun},\ and\
  \citenamefont {Lin}}]{su2022massive}%
  \BibitemOpen
  \bibfield  {author} {\bibinfo {author} {\bibfnamefont {Y.}~\bibnamefont
  {Su}}, \bibinfo {author} {\bibfnamefont {H.}~\bibnamefont {Li}}, \bibinfo
  {author} {\bibfnamefont {C.}~\bibnamefont {Zhang}}, \bibinfo {author}
  {\bibfnamefont {K.}~\bibnamefont {Sun}},\ and\ \bibinfo {author}
  {\bibfnamefont {S.-Z.}\ \bibnamefont {Lin}},\ }\bibfield  {title} {\bibinfo
  {title} {Massive dirac fermions in moir\'e superlattices: A route towards
  topological flat minibands and correlated topological insulators},\ }\href
  {https://doi.org/10.1103/PhysRevResearch.4.L032024} {\bibfield  {journal}
  {\bibinfo  {journal} {Phys. Rev. Res.}\ }\textbf {\bibinfo {volume} {4}},\
  \bibinfo {pages} {L032024} (\bibinfo {year} {2022})}\BibitemShut {NoStop}%
\bibitem [{\citenamefont {Bernevig}\ \emph {et~al.}(2006)\citenamefont
  {Bernevig}, \citenamefont {Hughes},\ and\ \citenamefont
  {Zhang}}]{bernevig2006quantum}%
  \BibitemOpen
  \bibfield  {author} {\bibinfo {author} {\bibfnamefont {B.~A.}\ \bibnamefont
  {Bernevig}}, \bibinfo {author} {\bibfnamefont {T.~L.}\ \bibnamefont
  {Hughes}},\ and\ \bibinfo {author} {\bibfnamefont {S.-C.}\ \bibnamefont
  {Zhang}},\ }\bibfield  {title} {\bibinfo {title} {{Quantum spin Hall effect
  and topological phase transition in HgTe quantum wells}},\ }\href@noop {}
  {\bibfield  {journal} {\bibinfo  {journal} {science}\ }\textbf {\bibinfo
  {volume} {314}},\ \bibinfo {pages} {1757} (\bibinfo {year}
  {2006})}\BibitemShut {NoStop}%
\bibitem [{\citenamefont {Zhao}\ \emph {et~al.}(2021)\citenamefont {Zhao},
  \citenamefont {Xiao},\ and\ \citenamefont {Yao}}]{zhao2021universal}%
  \BibitemOpen
  \bibfield  {author} {\bibinfo {author} {\bibfnamefont {P.}~\bibnamefont
  {Zhao}}, \bibinfo {author} {\bibfnamefont {C.}~\bibnamefont {Xiao}},\ and\
  \bibinfo {author} {\bibfnamefont {W.}~\bibnamefont {Yao}},\ }\bibfield
  {title} {\bibinfo {title} {{Universal superlattice potential for 2D materials
  from twisted interface inside h-BN substrate}},\ }\href@noop {} {\bibfield
  {journal} {\bibinfo  {journal} {npj 2D Materials and Applications}\ }\textbf
  {\bibinfo {volume} {5}},\ \bibinfo {pages} {38} (\bibinfo {year}
  {2021})}\BibitemShut {NoStop}%
\bibitem [{\citenamefont {Brzezi{\'n}ska}\ and\ \citenamefont
  {Yazyev}(2023)}]{brzezinska2023flat}%
  \BibitemOpen
  \bibfield  {author} {\bibinfo {author} {\bibfnamefont {M.}~\bibnamefont
  {Brzezi{\'n}ska}}\ and\ \bibinfo {author} {\bibfnamefont {O.~V.}\
  \bibnamefont {Yazyev}},\ }\bibfield  {title} {\bibinfo {title} {Flat bands in
  bilayer graphene induced by proximity with polar $ h $-{BN} superlattices},\
  }\href@noop {} {\bibfield  {journal} {\bibinfo  {journal} {arXiv preprint
  arXiv:2305.09749}\ } (\bibinfo {year} {2023})}\BibitemShut {NoStop}%
\bibitem [{\citenamefont {Angeli}\ and\ \citenamefont
  {MacDonald}(2021)}]{angeli2021gamma}%
  \BibitemOpen
  \bibfield  {author} {\bibinfo {author} {\bibfnamefont {M.}~\bibnamefont
  {Angeli}}\ and\ \bibinfo {author} {\bibfnamefont {A.~H.}\ \bibnamefont
  {MacDonald}},\ }\bibfield  {title} {\bibinfo {title} {{$\Gamma$ valley
  transition metal dichalcogenide moir{\'e} bands}},\ }\href@noop {} {\bibfield
   {journal} {\bibinfo  {journal} {Proceedings of the National Academy of
  Sciences}\ }\textbf {\bibinfo {volume} {118}},\ \bibinfo {pages}
  {e2021826118} (\bibinfo {year} {2021})}\BibitemShut {NoStop}%
\end{thebibliography}%

\end{document}